\documentclass[apj]{emulateapj}
\usepackage{natbib}

\usepackage{float} 

\usepackage{subfigure}
\usepackage{rotating} 
\usepackage{empheq}
\usepackage{verbatim}

\shorttitle{Stochastic acceleration and spectral curvature in blazars}
\shortauthors{A. Tramacere, E. Massaro \& A. M. Taylor}

\begin{document}

\title{Stochastic acceleration and the evolution of spectral distributions in SSC sources:
	A self consistent modeling of blazars' flares}

\author{A.~Tramacere}
\affil{
ISDC, University of Geneva, Chemin d'Ecogia 16, Versoix, CH-1290, Switzerland}
\email{andrea.tramacere@unige.ch}

\author{E.~Massaro} 
\affil{
Dipartimento di Fisica, Universit\`a La Sapienza, Piazzale A. Moro 2, 
I-00185 Roma, Italy}
    
\author{A.~M.~Taylor}
\affil{ISDC, University of Geneva, Chemin d'Ecogia 16, Versoix, CH-1290, Switzerland}



\begin{abstract}
The broad-band spectral distributions of non-thermal sources, such as
those of several known blazars, are well described by a log-parabolic fit.
The second degree term in these fits measures the curvature in the spectrum.  
In this paper we 
investigate whether the curvature parameter observed in the spectra of
the synchrotron emission can be used as a fingerprint of  stochastic 
acceleration.  

As a first  approach we use the multiplicative Central Limit theorem to show how
 fluctuations in the energy gain result in the
broadening of the spectral shape, introducing a curvature into the energy
distribution.  Then, by means of a Monte-Carlo description, we investigate how the 
curvature produced in the electron  distribution is linked to the diffusion
in  momentum space. 
To get a more generic  description of the
problem we turn to the diffusion equation in  momentum space. We first
study some ``standard'' scenarios, in order to understand 
the conditions that make the curvature in the spectra significant, and 
the relevance of cooling during the acceleration process. 
We try to quantify the correlation between the curvature and the
diffusive process in the pre-equilibrium stage, and investigate how the transition 
between the Klein-Nishina  and the Thompson regime, in Inverse Compton cooling,  
determine the curvature in the distribution at  equilibrium.
We apply these results to some observed trends, such as
the anticorrelation between the peak energy and the curvature term
observed in the spectra of Mrk 421, and a sample of BL Lac objects whose 
synchrotron emission peaks at X-ray energies
\end{abstract}

\keywords{acceleration mechanisms: statistical - galaxies: active - galaxies: 
BL Lacertae objects - galaxies:  BL Lacertae objects (Mrk 421, Mrk 501, 1H 1426+428, 
1ES 1959+650, Mrk 180, PKS 0548-322) }





\section{Introduction}
\label{sec:Intro}

A defining feature of the non-thermal emission from different types of
galactic and extragalactic sources is that their spectra are described
by a power-law (PL) over a broad photon energy range.  In several sources,
however, their spectra show  significant curvature that is typically milder
than that expected from an exponential cut-off.  In previous
papers \cite{Massaro2004,Massaro2006}  discussed the
curvature observed in the broad band X-ray spectra of the two well
known HBL (High-energy peaked BL Lac) objects Mkn~421 and Mkn~501.
The basic idea was that this curvature was not simply the result of
radiative cooling of high energy electrons, responsible of the
synchrotron and inverse Compton emission, but that it was essentially
related to the acceleration mechanism.
\cite{Massaro2004}  showed that  curved spectral distributions, in particular  
log-parabolic (i.e. log-normal) ones, develop when the acceleration 
probability is a decreasing function of the electron energy.  
In subsequent works, through the analysis of a large collection of X-ray
observations of Mkn 421, 
\cite{Tramacere2007Mrk421,TramacerePhD2007,Tramacere2009} pointed out that 
the observed anticorrelation between the peak energy and the curvature measured 
in the synchrotron Spectral Energy Distribution (SED), could be used as a clear 
signature of a stochastic component in the acceleration process. 
Very recently, the log-parabolic law has been also applied to describe the
spectral distribution and evolution of some Gamma-ray bursts \citep{Massaro2011}.

The principal aim of the present paper is to investigate this scenario by 
extracting information on the acceleration processes using the curvature parameter 
measured in the observed synchrotron and Inverse  Compton (IC) spectra of
Synchro-Self Compton (SSC) sources.    
We study the conditions in which the energy distributions of
electrons, resulting from stochastic acceleration, can be approximated
by a log-parabolic law and how its curvature evolves during their
acceleration, and the role of IC cooling. 
We compare predictions from our theoretical descriptions with the curved spectra 
of some HBL objects.

In Sec. \ref{sec:Logpar} we give an intuitive picture to take into account the effect  
of random fluctuations in the energy gain of particles and the role these
play in determining the spectral curvature, as a consequence of the multiplicative 
central limit theorem,  and compare these results with the analytical solution 
of the diffusion equation, in the ``hard-spheres" approximation.
In Sec. \ref{sec:MC} and Sec. \ref{sec:numeric} we give a more physical 
description of the problem, using first a Monte Carlo approach, and secondly by 
solving numerically the momentum diffusion equation. We discuss the 
evolution of the curvature in the electron distribution as a result of 
momentum-diffusion before equilibrium is reached, 
and the role that synchrotron and IC cooling processes play on reaching
the equilibrium. In Sec. \ref{sec:ssc_evolution},  we study the peak energy, 
fluxes, and curvature, trends in the SED of both the synchrotron and IC emission, 
looking for the fingerprints of the stochastic component. 
In Sec. \ref{sec:fit_data}, we show how our results can reproduce the spectral
trends observed in a some HBLs, in particular we investigate the relation
between the peak energy and the curvature, and between the peak energy and the
peak flux. The good agreement between predictions and observed trends, confirms
that the stochastic acceleration mechanism can play an important role in the
physics of the blazars' jets and other SSC sources.


\section{The Log Parabola Origin: analytical approach}
\label{sec:Logpar}
\subsection{Statistical description}
\label{sec:StatApproach}
In the statistical picture, the change in energy of the 
particles at each acceleration step $n_s$ is expressed as
\begin{equation}    
 \gamma_{n_s} =\varepsilon_{n_s} \gamma_{n_s-1} = \gamma_{n_s-1}(1+\Delta \gamma_{n_s-1}/\gamma_{n_s-1})
 \label{eq:ene_acc}
\end{equation}
where $\gamma$ is the Lorentz factor of the particle and $\varepsilon$ is the 
fractional energy gain. We here investigate the role of fluctuations of $\varepsilon$,
on the spectral shape of the accelerated particles.  
With this aim in mind, we express the energy gain fluctuations as
\begin{equation}  %
\varepsilon = \bar{\varepsilon} +\chi
 \label{eq:ene_gain}
\end{equation}
where the random variable $\chi$ has a probability density function
with zero mean  value ($\langle \chi\rangle=0$) and variance 
$\sigma_{\chi}^2$, and $\bar{\varepsilon}$ represents the systematic
energy gain, that we treat as a non-random variable and the probability density 
function of $\varepsilon$ is defined on the range  $\varepsilon\geq0$.
The particle energy at step $n_s$ can be
expressed as:
\begin{equation}    
 \gamma_{n_s} =\gamma_{0} \Pi_{i=1}^{n_s }\varepsilon_{i}
 \label{eq:ene_acc_0}
\end{equation}
where $\gamma_0$ is the initial energy of the particle. This equation 
clearly shows that  the final energy distribution ($n(\gamma)=dN(\gamma)/d\gamma$) 
will result from the product of the  random variables $\varepsilon_i$.
The determination of an analytic expression  for the distribution resulting from the 
multiplication of generic random variable is not an easy task (Glen et al. 2004).
Using the simplifying assumption that the particles are always 
accelerated, namely the acceleration probability, $P_a$, is set to unity and applying the 
multiplicative case of the central limit theorem (e.g. Cowan 1998) it is possible
to show that the particle energies will be distributed as a log-normal law:
\begin{equation}   
n(\gamma)=\frac{N_0}{\gamma\sigma_{\gamma}  \sqrt(2\pi)} \exp\Big[-(\ln~\gamma -
\mu)^2/2\sigma_{\gamma}^2\Big]
\label{eq:log_norm_0}
\end{equation}
where $N_0$, is the total number of particles, $\mu=\langle\rm{ln}~ \gamma\rangle$, 
$\sigma_{\gamma}^2=\sigma^2(\rm{ln}~ \gamma)$.
We can determine these two quantities by taking the logarithm of Eq. \ref{eq:ene_acc_0},
\begin{eqnarray}    
\ln  ~\gamma_{n_s} &=&\ln~\gamma_{0} + \Sigma_{i=1}^{n_s} \ln~(\bar\varepsilon+ \chi_i) \nonumber\\
&=&   \ln~(\gamma_{0}\bar\varepsilon^{n_s})+ \Sigma_{i=1}^{n_s} \ln \Big( 1 +\frac{\chi_i}{\bar\varepsilon} \Big) \nonumber \\
&\approx&\ln~(\gamma_{0}\bar{\varepsilon}^{n_s})+\Sigma_{i=1}^{n_s} \Big( \frac{\chi_i}{\bar{\varepsilon}}- \frac{\chi_i^2}{2\bar{\varepsilon}^2 } \Big)
\label{eq:ene_acc_1}
\end{eqnarray}
assuming that $\chi_i/\bar{\varepsilon}$ is not large. 
We obtain for the two parameters $\mu$ and $\sigma_{\gamma}$:
\begin{eqnarray}   
\mu&=&\rm{ln}~(\gamma_{0}) +n_s~\ln~ \bar{\varepsilon} ~+~ n_s\Big[
\langle\frac{\chi}{\bar{\varepsilon}}\rangle-
\frac{1}{2}\Big(\frac{\sigma_{\chi}}{\bar{\varepsilon}}\Big)^2
-\langle\frac{\chi}{2\bar{\varepsilon}}\rangle^2
\Big]\nonumber \\
\sigma_{\gamma}^2 &=& \rm{n_s}
\Big[
\Big(\frac{\sigma_{\chi}}{\bar{\varepsilon}}\Big)^2+ 
\Big(\frac{\sigma_{\chi}}{2\bar{\varepsilon}}\Big)^4+
2\Big(\frac{\sigma_{\chi}}{2\bar{\varepsilon}}\langle\frac{\chi}{2\bar{\varepsilon}}\rangle \Big)^2
\Big]
\label{eq:sigma_gamma_0}
\end{eqnarray}
where we have ignored the covariance terms since we are assuming 
the energy gain at each acceleration step being independent on the one at the 
previous step.
Remembering that $\langle \chi\rangle=0$, $\sigma_{\chi}=\sigma_{\varepsilon}$,  
and ignoring the 4-th order term, we can write:
\begin{eqnarray}   
\mu&=&\ln~(\gamma_{0})  ~+~ n_s\Big[
\ln~ \bar{\varepsilon}-
\frac{1}{2}\Big(\frac{\sigma_{\varepsilon}}{\bar{\varepsilon}}\Big)^2
\Big] \\ \nonumber
\sigma_{\gamma}^2 &\approx& n_s\Big(\frac{\sigma_{\varepsilon}}{\bar{\varepsilon}}\Big)^2
\label{eq:sigma_gamma}
\end{eqnarray}

This equation shows that the variance increases linearly with the number of acceleration steps 
and it is proportional  to ${\sigma_{\varepsilon}}^2$. 
Substituting  $\mu$ and $\sigma_{\gamma}$  into Eq. \ref{eq:log_norm_0},
\begin{equation}   
n(\gamma)=\frac{N_0}{\gamma\sigma_{\gamma}  \sqrt(2\pi)} 
\exp\Big[
\frac{-\Big(\ln \frac{\gamma}{\gamma_0} - n_s\Big[\ln~ \bar{\varepsilon}-
\frac{1}{2}\Big(\frac{\sigma_{\varepsilon}}{\bar{\varepsilon}}\Big)^2
\Big]\Big)^2}{2n_s\Big(\frac{\sigma_{\varepsilon}}{\bar{\varepsilon}}\Big)^2}
\Big]
\label{eq:log_norm}
\end{equation}

Hereafter we will consider decimal logarithms
(log $\equiv$ log$_{10}$, $c_e=1/\log_{10} e ~\approx 2.3$), to make easier a
comparison of the curvature results     form this paper with those presented in
observational papers.  
Taking the logarithm of Eq. \ref{eq:log_norm}, and substituting the parameters from 
Eq. \ref{eq:sigma_gamma} we obtain:
\begin{equation}  
\log n(\gamma) = K - \log \gamma - 
\frac{\Big(c_e\log \frac{\gamma}{\gamma_0} - n_s\Big[c_e\log~ \bar{\varepsilon}- 
\frac{1}{2}\Big(\frac{\sigma_{\varepsilon}}{\bar{\varepsilon}}\Big)^2
\Big]\Big)^2}{c_e2n_s\Big(\frac{\sigma_{\varepsilon}}{\bar{\varepsilon}}\Big)^2}
\label{eq:log_par}
\end{equation}    
where $K$ includes all the constant factors. 
This is a log-parabolic law with the curvature (2nd degree in $\log~ \gamma$) 
coefficient given by:
\begin{equation}   
r=\frac{c_e}{2n_s\Big(\frac{\sigma_{\varepsilon}}{\bar{\varepsilon}}\Big)^2}~~~.
\label{eq:curvature_ene_disp}
\end{equation}
The interesting physical insight of this equation is that the curvature
of the particle energy distribution is inversely proportional to
the acceleration steps ($n_s$),
and to the variance of the energy gain ($\sigma_{\varepsilon}^2$).
In the case of $P_a<1$, the distribution at step $n_s$ will be given
by the convolution of different log-normal distributions for each
acceleration step, with the distribution at $n_s$ broader
than that at $n_{s}-1$  and containing fewer particles, as already noted in  
\cite{Peacock1981}.

Similar results are obtained considering a constant energy gain but a
fluctuating number of acceleration steps. 
Assuming that after a time $t$ the probability distribution for the number
of steps undergone by a particle is given by a Poisson law, it is possible
to show that the energy distribution follows a log-parabola whose
curvature term depends on the inverse of the mean number of steps multiplied
by the duration of the acceleration process.

\subsection{Diffusion equation approach} 
\label{sec:DiffEqApproach}
The above statistical description provides an intuitive link between the curvature in
the energy distribution of accelerated particles and the presence of a
randomization process, such as the dispersion in the energy gain or in the number of
acceleration steps. However, this approach does not give a complete physical
description of the processes responsible for the systematic and stochastic
energy gain, ignoring other physical processes, such as the radiative cooling and 
injection rates, or the acceleration energy dependence, necessary to give a complete description of the particles
energy distribution evolution.
A physical self-consistent description of stochastic acceleration in a
time-dependent fashion, can be achieved through a kinetic equation approach. 
Employing the quasi-linear approximation with the inclusion of momentum diffusion term 
\citep{Ramaty1979,Becker2006}, the equation governing the temporal evolution
of $n(\gamma)$ is: 
\begin{eqnarray}
\frac{\partial n(\gamma,t)}{\partial t}&=&
\frac{\partial }{\partial \gamma }
\Big\{ 
-[S(\gamma,t) + D_A(\gamma,t) ]n(\gamma,t)\Big\}\\
&+&\frac{\partial }{\partial \gamma }
\Big\{ 
D_{p}(\gamma,t) \frac{\partial n(\gamma,t)}{\partial \gamma }\nonumber
\Big\}
-\frac{n(\gamma,t)}{T_{esc}(\gamma)}+Q(\gamma,t)
\label{eq:diff_eq}
\end{eqnarray}
where $D_{p}(\gamma,t)$ is the momentum diffusion coefficient,
$D_A(\gamma,t)=(2/\gamma)D_{p}(\gamma,t)$ is the average energy change term
resulting from the momentum-diffusion process, and
$S(\gamma,t)=-C(\gamma,t)+A(\gamma,t)$ is an extra term describing systematic 
energy loss ($C$) and/or gain ($A$), and $Q(\gamma,t)$ is the injection term. In
the standard diffusive shock acceleration scenario, there are several
possibilities for which one can expect that energy gain fluctuations will occur,
due to the momentum diffusion term. In particular,  for the case of a turbulent
magnetized medium, the advection of particles towards the shock 
due to pitch angle scattering may be accompanied by  stochastic momentum
diffusion mechanism. In this scenario, particles embedded in a magnetic field
with both an ordered ($B_0$) and turbulent ($\delta B$) component, exchange energy
with resonant plasma waves, and  the related diffusion  coefficient is
determined by the spectrum of the plasma waves. Following the approach of
\cite{Becker2006} we describe the energy distribution $W(k)$ in terms of the
wave number $k=2\pi/\lambda$ with a power-law :
\begin{eqnarray}
W(k)=\frac{\delta B(k)^{2}}{8\pi}=\frac{\delta B(k_{0})^{2}}{8\pi}\left(\frac{k}{k_{0}}\right)^{-q}.
\label{eq:turb_spec}
\end{eqnarray}
with $q=2$ for the ``hard-sphere'' spectrum, $q=5/3$ for the Kolmogorov spectrum, 
and $q=3/2$ for the Kraichnan spectrum, 
the total energy density in the fluctuations being 
\begin{eqnarray}
U_{\delta B}=\int_{k_{0}}^{k_{\rm max}}W(k)dk~~~.
\end{eqnarray}
Under these assumptions the momentum-diffusion coefficient reads \citep{OSullivan:2009p1344}:
\begin{eqnarray}
\label{eq:diff_coeff_0}
D_{p}\approx\beta_{A}^2\Big(\frac{\delta B}{B_0}\Big)^2   \Big(\frac{\rho_g}{\lambda_{max}}\Big)^{q-1}\frac{p^2c^2}{\rho_g c}
\end{eqnarray}
where $\beta_{A}=V_A/c$ and $V_A$ is the 
Alfven waves velocity, $\rho_g=pc/qB$ is the Larmor radius, and $\lambda_{max}$ is the maximum wavelength
of the Alfven waves spectrum.
The acceleration time for particles with Lorentz factor $\gamma$, whose Larmor radii
resonate with one particular magnetic field turbulence length-scale, is dictated 
by the momentum diffusion coefficient ($D_p$) as,
\begin{eqnarray}
t_{\rm acc}\approx \frac{p^{2}}{D_{p}}=
\frac{\rho_{g}(\gamma_{0})}{c~\beta_{A}^{2}}
\left.\left(\frac{B_{0}^{2}}{\delta B^{2}}\right)\right|_{\gamma_{0}}\left(\frac{\gamma}{\gamma_{0}}\right)^{2-q}.
\label{eq:tacc}
\end{eqnarray} 
The spatial diffusion coefficient relates to the momentum diffusion coefficient 
through the relation, $D_{x}D_{p}\approx p^{2}\beta_{\rm A}^{2}$ \citep{skilling1975},
hence  the escape time of the particles from the acceleration region
of size $R$, depends on the spatial diffusion coefficient through the relation,
\begin{eqnarray}
t_{\rm esc}\approx \frac{R^{2}}{D_{x}}\approx \frac{R^{2}}{\left(c\beta_{\rm A}\right)^{2}t_{\rm acc}}.
\label{eq:tesc}
\end{eqnarray}
The coefficients in Eq. \ref{eq:diff_eq},  and their related  time scales, can be 
expressed as a power-law in terms of the Lorentz factor ($\gamma$):
\begin{equation}
\begin{cases} 
D_{p}(\gamma)&=D_{p0}\left(\frac{\gamma}{\gamma_{0}}\right)^{q},~~~~~~~~ t_D =\frac{1}{D_{p0}}\left(\frac{\gamma}{\gamma_{0}}\right)^{2-q} \\
D_A(\gamma)&=2D_{p0}\left(\frac{\gamma}{\gamma_{0}}\right)^{q-1},~~~t_{DA} = \frac{1}{2D_{p0}}\left(\frac{\gamma}{\gamma_{0}}\right)^{2-q}\\
A(\gamma)&=A_{p0}\gamma,~~~~~~~~~~~~~~~~ t_{A} = \frac{1}{A_{0}}\\
\end{cases} 
\label{eq:diff_coeff}
\end{equation}
where $D_{p0}$, and $A_0$ have the dimension of the inverse of a time. 
Analytical solutions of the diffusion equation for relativistic electrons
are frequently discussed in the literature since the early work by
\cite{Kardashev1962},  in particular for the case of the  ``hard-sphere'' approximation. 
Neglecting  the $S$ and $T_{esc}$ terms in Eq. \ref{eq:diff_eq}, and using  
a mono-energetic and instantaneous 
injection ($n(\gamma,0)=N_0\delta(\gamma-\gamma_0)$), the solution of the diffusion 
equation is \citep{Melrose1969,Kardashev1962}:
\begin{equation}
 n(\gamma,t)=
\frac{N_0}{\gamma\sqrt{4\pi D_{p0} t}}
\exp{ \Big\{
-\frac{ [\ln(\gamma/\gamma_0)-(A_{p0} - D_{p0}) t ]^2 }{4D_{p0}t}   
\Big\},
}
\label{eq:analyt_sol}
\end{equation}
ie. a log-parabolic distribution, whose curvature term is:
\begin{equation}
 r=\frac{c_e}{4D_{p0}~t~} \propto \frac{1}{D_{p0}t}
 \label{eq:analyt_sol_curv}
\end{equation}
This result is fully consistent with that found  in the statistical
description, indeed Eq. \ref{eq:analyt_sol} and Eq. \ref{eq:log_norm} 
have the same functional form in both the statistical and  in the diffusion equation
scenario, with $t$ playing the role of $n_{\rm s}$, $D_{p0}$ the role of the 
variance of the energy gain ($\sigma_\varepsilon^2$), and $A_{p0}$ the role of
$\log~ \bar\varepsilon$. Hence we can write: 
\begin{equation}
D_{p0}\propto \Big(\frac{\sigma_\varepsilon}{\bar\varepsilon}\Big)^2
\label{eq:dp_vs_sigma}
\end{equation}
It is interesting to note, that in the case of the 
``hard-sphere'' approximation, the curvature term is simply dictated by the ratio of the
diffusive acceleration time ($t_{D}$) to the evolution time ($t$).
\\
\\

\section{Numerical approach: Monte-Carlo simulation with magnetic turbulence}

\label{sec:MC}
In this section we demonstrate explicitly how the introduction of energy 
fluctuations leads to curved spectral distributions of particles. 
This is carried out using a Monte-Carlo (MC) approach.

In our simulations, we considered $10^{5}$ particles injected into the system with
a cold mono-energetic distribution of Lorentz factors, with $\gamma_0=1$. 
To compare these results with the ones presented in Sec. \ref{sec:Logpar}, we remind 
the reader that, 
in the MC approach, the duration of the acceleration process $t$ is the equivalent 
of the number of acceleration step ($n_s$) used in the statistical picture,
and that the probability of the particle to be up-scattered or down-scattered in the MC
realizations, can be expressed in the statistical approach as  $P(\varepsilon>1)$ 
and $P(\varepsilon<1)$, respectively.
The scattering probability of the particles is dictated by
the intensity of resonant waves in the turbulent magnetic power spectrum.
As a working hypothesis we assume that particles interact with a turbulent 
magnetic field whose power spectrum is expressed by Eq. \ref{eq:turb_spec}.
In each scattering, the particles have probability $(1+\beta_{\rm A})/2$
of being up-scattered, and probability $(1-\beta_{\rm A})/2$ of being down-scattered.
The energy dispersion of the particle due to resonant scattering with 
Alfven waves will be $\langle\Delta E^2\rangle\propto (E\beta_{\rm A})^2t$,
where $E=m_ec^2\gamma$. Using the very good approximation for the variance
of the product of $n$ uncorrelated random variables \citep{Goodman1962}:
\begin{equation}
\sigma^2(\Pi x_i)=\Pi\langle x_i \rangle^2 \Sigma ~\left(\frac{\sigma_{x_{i}}^2}{\langle x_i \rangle^2} \right)
    \end{equation}
and plugging Eq. \ref{eq:ene_gain} into the equation  above, we get:
\begin{equation}
\langle\Delta E^2\rangle\propto (E\beta_{\rm A})^2t\propto\gamma_0^2\sigma^2(\Pi\varepsilon_i )=(\gamma_0\bar\varepsilon^{n_s})^2~n_s~\frac{\sigma_\varepsilon^2}{\bar\varepsilon^2}
\end{equation}
since $E$ is the particle energy at time $t$ (namely step $n_s$), we have
$E^2=(m_ec^2\gamma_0\bar\varepsilon^{n_s})^2$, from which follows:

\begin{equation}
\beta_{\rm A}^2\propto
\Big(\frac{\sigma_\varepsilon}{\bar\varepsilon}\Big)^2
\label{eq:beta_vs_sigma}
\end{equation}


\begin{figure*}[ht]
\centering
\begin{tabular}{ll}
\includegraphics[width=0.45\linewidth,angle=-0]{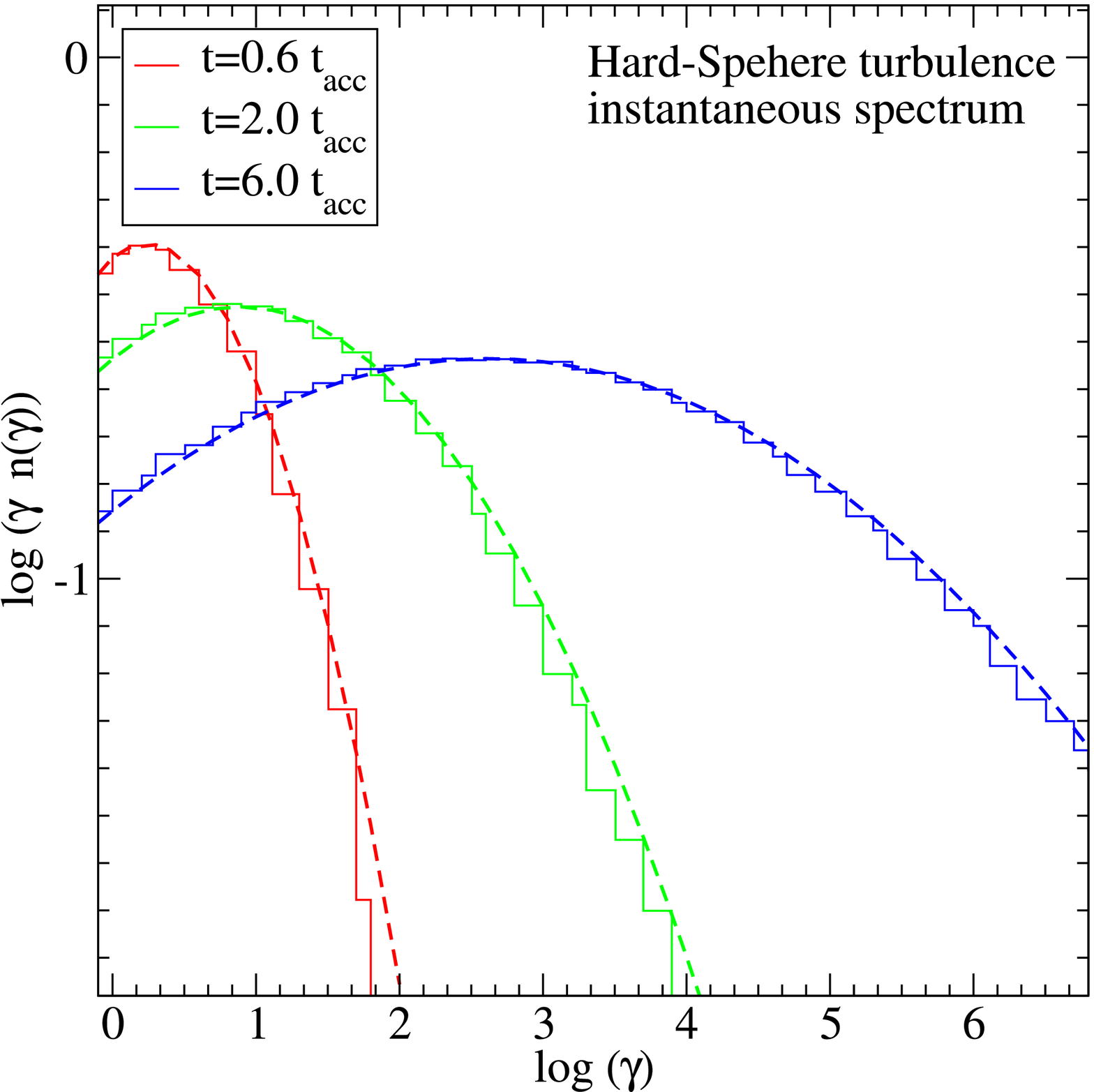}&
\includegraphics[width=0.45\linewidth,angle=-0]{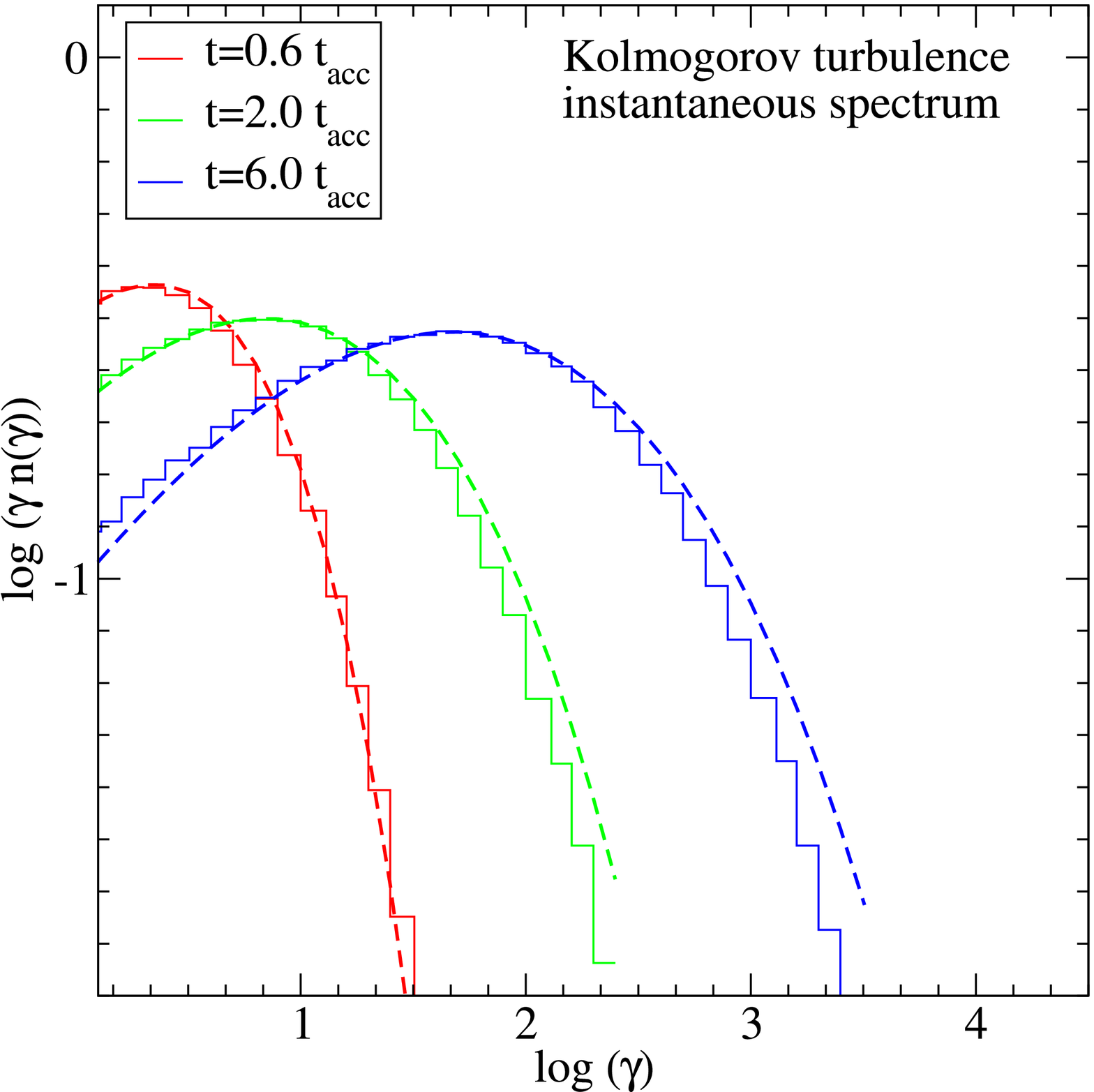}\\
\end{tabular}
\caption{Plots showing the Monte Carlo results. For comparison, the results obtained using 
an analytic description given in \cite{Becker2006} (dashed line, right panel) and a  
log-parabolic function (dashed line, left panel), are shown.}
\label{fig:hard_Kolmog}
\end{figure*}   


In the following two sections (Sec. \ref{sec:HS_turb} and, Sec \ref{sec:soft_tur}) we 
study the consequences of the structure in the magnetic turbulence, on the evolution of the particle spectra,
following their stochastic acceleration in the turbulent field.


\subsection{Hard-Sphere Turbulence}
\label{sec:HS_turb}
Under the ``hard-sphere'' approximation ($q=2$), the spatial diffusion coefficient 
does not depend on the particle energy, since the exponent of Eq. \ref{eq:tacc} is $q-2=0$.
Only three independent parameters exist in this description: the scattering time,
the escape time and the velocity of the scatterers.
The spectra are purely determined by how many scatterings  
have been able to occur, the velocity of the scatterer, and what fraction 
of the injected particles have escaped out of the acceleration region.
The scattering time relates to the spatial diffusion coefficient by 
$t_{\rm scat}\approx D_{x}/c$.
Similarly, the resulting acceleration time relates to the spatial diffusion 
coefficient by 
$t_{\rm acc}\approx D_{x}/\beta_{\rm A}^{2}c\approx t_{\rm A}/\beta_{\rm A}^{2}$. 
Thus, for ``hard sphere'' turbulence, the scattering and acceleration time scales
are independent of the particle energy (since there is equal energy density
of scatterers which particles of all energies may resonantly scatter with).

\begin{figure}[!t]
\centering
\includegraphics[width=1.0\linewidth,angle=-0]{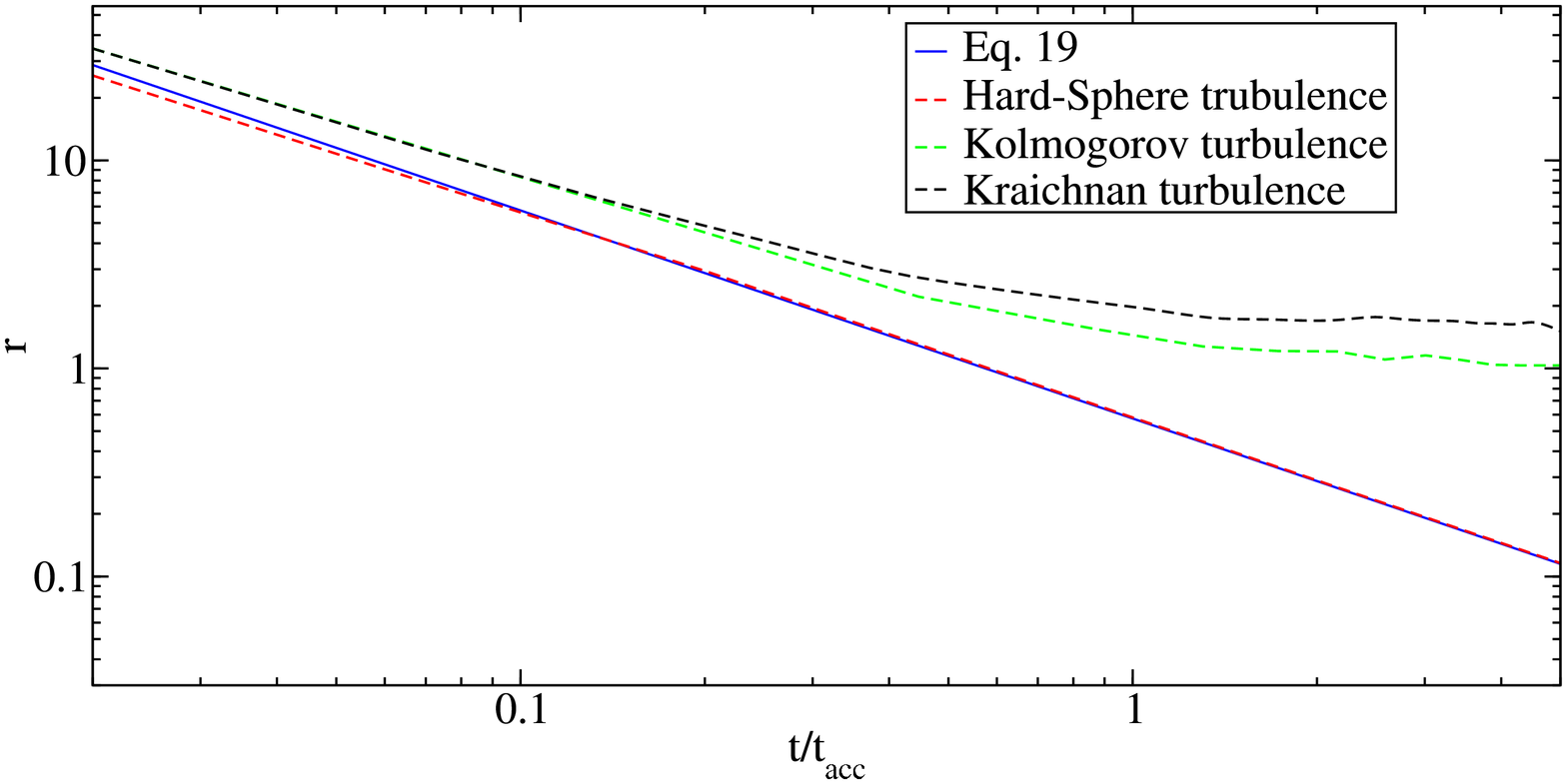}
\caption{
The curvature parameters of the energy distribution of accelerated 
electrons in the shown in Fig \ref{fig:hard_Kolmog}.
In the case of $q=2$ (red line), the trend is consistent with the "hard-spheres" prediction (blue line). 
In the case of Kolmogorov (green line) and Kraichnan (black line) turbulence, the trend predicts larger 
values compared to the "hard-spheres" prediction, and $r$ approaches an asymptotic value 
dictated by the exponential cut-off in the equilibrium distribution.
}
\label{fig:curv_MC}
\end{figure}

The left-hand panel  in Fig. \ref{fig:hard_Kolmog} shows the 
resulting instantaneous  evolution of spectra for the 
``hard sphere'' turbulence.
The log-parabolic shape is maintained along the entire acceleration process, 
as shown by the solid lines representing the fit of the MC distributions 
by means of the law in Eq. \ref{eq:log_par}.
The evolution of the curvature parameter, obtained from the $r$ in the log-parabolic fit, 
and plotted in Fig. \ref{fig:curv_MC} with the red dashed line, clearly shows the trend 
due to the momentum diffusion, in agreement with the prediction 
from Eq. \ref{eq:analyt_sol_curv} (blue line in the plot)
demonstrating the connection between $D_{p0}$, $\frac{\sigma_\varepsilon}{\bar\varepsilon}$, 
and $\beta_{\rm A}$.

\begin{figure*}[ht]
\centering
\begin{tabular}{cc}
\includegraphics[width=0.5\linewidth,angle=-0]{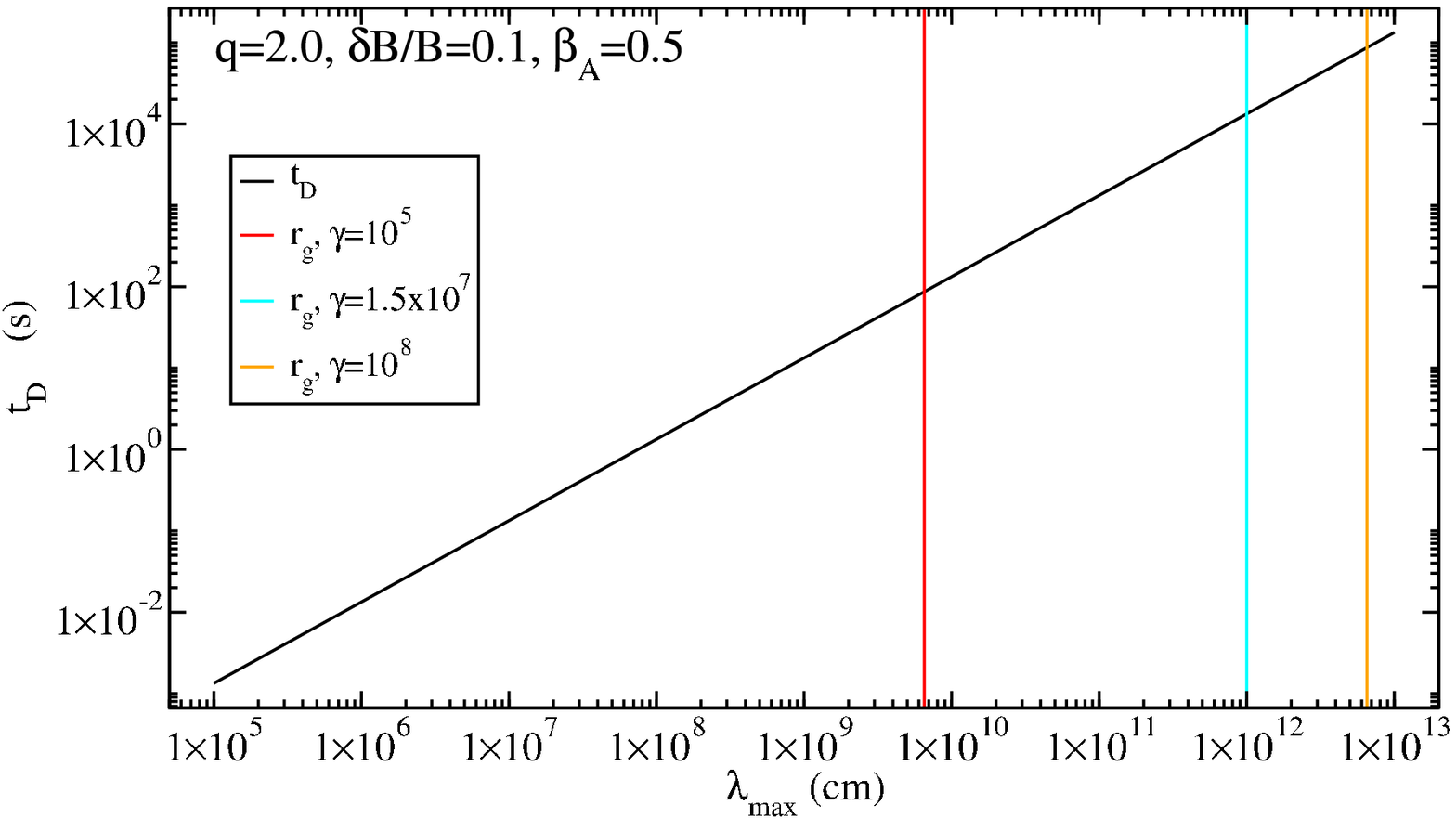}&
\includegraphics[width=0.5\linewidth,angle=-0]{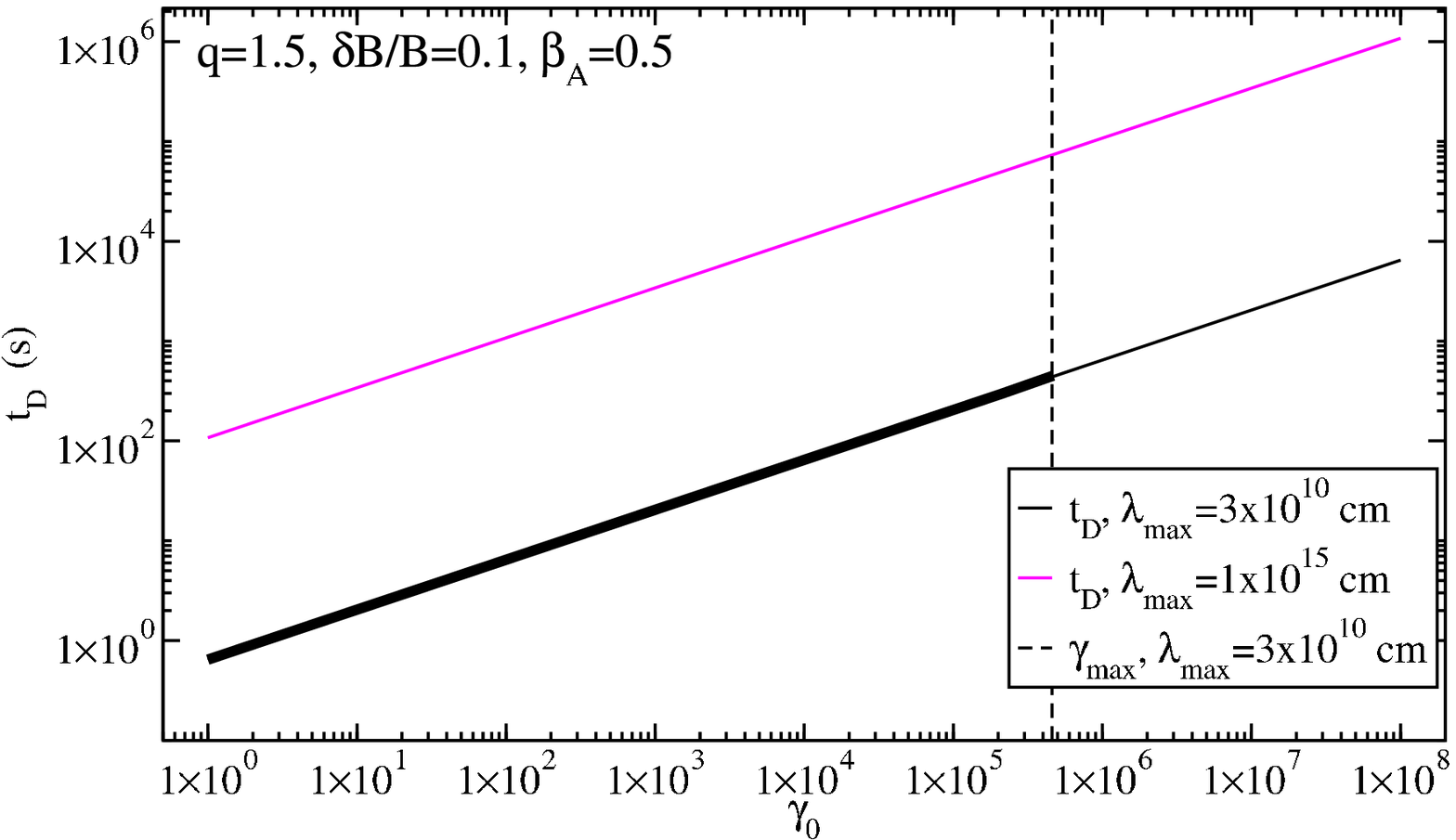}\\
\end{tabular}
\caption{\textit{Left panel:} the $t_D$ acceleration time as a 
function of $\lambda_{max}$, for $q=2$, $\delta B/B=0.1$, and $\beta_{A}=0.5$. 
The vertical lines represent the Larmor radius for $\gamma=10^5$ (red line),
$\gamma=1.5\times 10^{7}$ (cyan line), and $\gamma=10^8$ (orange line).
\textit{Right panel:} the $t_D$ acceleration time for the same parameters
as in right panel, for the case of $q=3/2$ and as  function of $\gamma$, 
for the two different cases of $\lambda_{max}=3\times 10^{10}$ cm (black line),
and $\lambda_{max}=1\times 10^{15}$ cm  (purple line). The thick black line shows $t_D$, 
for the case of $\lambda_{max}= 3\times 10^{10}$ cm, limited to the highest 
acceleration energy of the particles constrained by the resonant scattering limit: $\rho_g=\lambda_{max}$.}
\label{fig:t_D}
\end{figure*}

\subsection{Soft turbulence spectra}
\label{sec:soft_tur}
To account for the effects of turbulent magnetic field spectra softer 
than the ``hard-sphere'' case, we also consider acceleration in  Kolmogorov
and Kraichnan type turbulence.
We have therefore to include a fourth parameter in the MC simulation, in 
addition to the three considered above: the turbulent field spectral slope 
$q$ (see Eq. \ref{eq:turb_spec}). 

The right-hand plot in Fig.~\ref{fig:hard_Kolmog} shows the evolution of spectra 
for the ``Kolmogorov'' turbulence case.
Similar spectra were obtained by \cite{Lemoine2003},
and \cite{OSullivan:2009p1344}, who integrated the trajectories of charged particles in a turbulent
magnetic field embedded in a fluid.
The results are compared to the  quasi-linear theory results, \cite{Becker2006}
(solid lines in Fig. \ref{fig:hard_Kolmog}).
We can identify two phases in the  temporal evolution. In the first phase, the 
spectral energy distributions are more symmetric, and the curvature evolves as 
in the $q=2$ case, while in the second phase they develop a low-energy power-law tail.
Fig. \ref{fig:curv_MC} shows that, for the Kolmogorov (green line) and the Kraichnan (black line) turbulence, 
$r$ is systematically larger compared to the "hard-sphere" case (red line),  
and that for $t \gtrsim 2\times t_{acc}$, $r$ approaches to an asymptotic value ($r\approx 1.2$ and $r
\approx 1.5$, for $q=5/3$ and $q=3/2$ respectively) ruled by the 
exponential cut-off in the equilibrium distribution.

\begin{figure*}[!t]
\centering
\begin{tabular}{l}
\includegraphics[width=18cm,angle=-0]{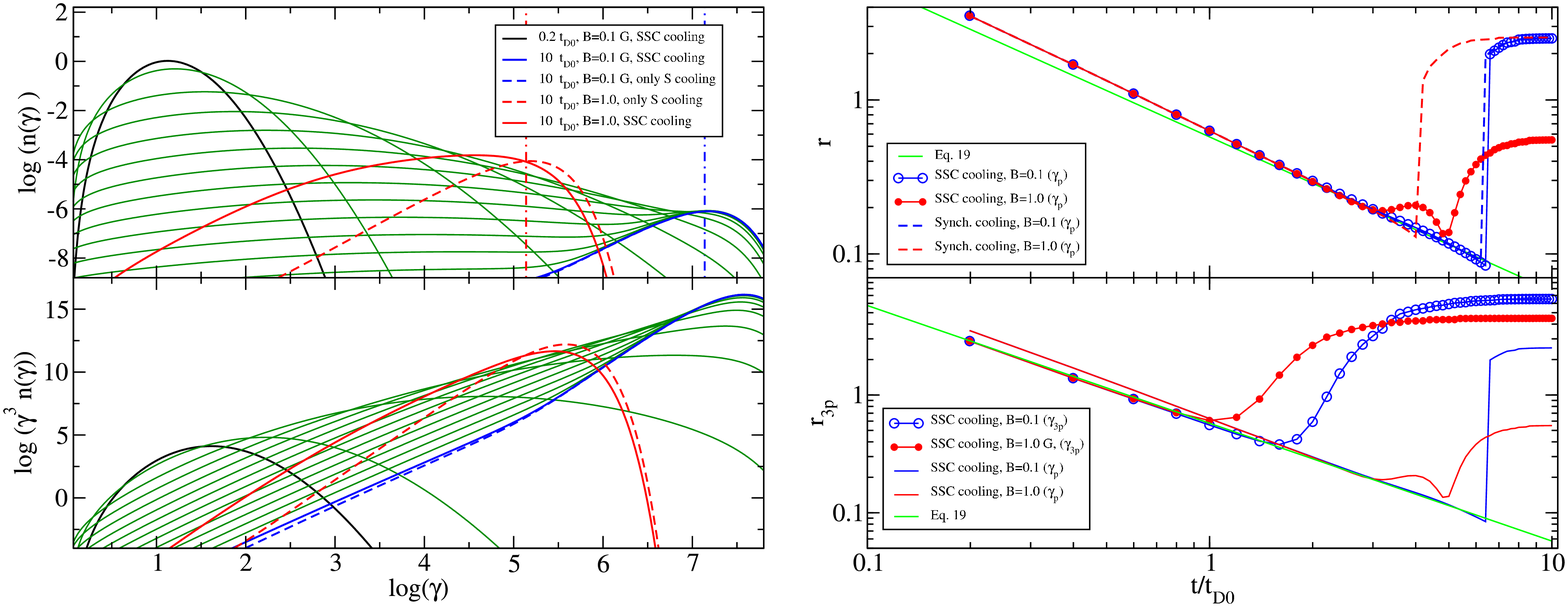} \\

\end{tabular}
\caption{ 
\textit{Left panels: } 
evolution of the particle spectrum with impulsive injection and no escape 
for the case of $R=1\times 10^{15} cm$ and $q=2$.
Upper panels represent the  temporal evolution of $n(\gamma)$, lower panels 
represent the temporal evolution of $\gamma^{3}n(\gamma)$.
Solid lines represent the case of SSC cooling. Red and blue solid lines, 
represent the final state for  $B=1.0$ G and $B=0.1$ G, respectively. 
Green solid lines represent the temporal evolution, for $B=0.1$ G, with step of $0.8\times t_D$.
The dashed lines represent the final stage in the case of only synchrotron cooling.
The vertical dot-dashed lines represent the equilibrium energy in the case
on only synchrotron cooling.
\textit{Right panels:} 
Evolution of the curvature as function of $t/t_{D_0}$. Upper panel: curvature $r$
evaluated at $\gamma_p$, for the case of SSC cooling (solid red and blue lines)
and for the case of only synchrotron cooling (dashed red and blue lines).
The solid green line represent the prediction from Eq. \ref{eq:analyt_sol_curv}.
Lower panel: the same as in the upper panel, for the curvature $r_{3p}$ 
evaluated at $\gamma_{3p}$ (empty and filled circles) compared to the case
of $r$ (solid lines).
}
\label{fig:n_evola}
\end{figure*}

\section{Numerical approach: diffusion equation with stochastic component
and losses} 
\label{sec:numeric}
Both MC approach and statistical description are  able to explain the  link between the 
curvature in the energy distribution of accelerated particles and the presence 
of a stochastic energy gain term. In order to incorporate a more complete 
description, taking into account the competition between radiative losses 
and acceleration, and its influence on the curvature, we use 
the diffusion equation approach, already
outlined in Sec \ref{sec:DiffEqApproach}, by inserting into Eq. \ref{eq:diff_eq} 
a cooling term for the synchrotron and IC radiative losses.
Following \cite{Moder2001} we can write:
\begin{eqnarray}
|\dot\gamma_{synch}|&=&\frac{ 4\sigma_Tc}{3 m_ec^2}\gamma^2 U_B=C_0\gamma^2U_B \\
|\dot\gamma_{IC}|&=&\frac{ 4\sigma_Tc}{3 m_ec^2}
\gamma^2 \int f_{KN}(4\gamma\epsilon_0)\epsilon_0 n_{ph}(\epsilon_0)d\epsilon_0=
C_0\gamma^2F_{KN}(\gamma)\nonumber \\
C(\gamma) &=&|\dot\gamma_{synch}|+|\dot\gamma_{IC}|  =C_0\gamma^2 ( U_B+  F_{KN}(\gamma)) \nonumber
\label{eq:cooling}
\end{eqnarray}
where $U_B=B^2/8\pi$, is the energy density of the magnetic field, $\epsilon_0=h
\nu_0/m_ec^2$ is the IC seed photon energy in units of $m_ec^2$, $n_{ph}(\epsilon_  0)$ is
the number density of IC seed photons with the corresponding photon energy density
$U_{ph}=m_ec^2\int \epsilon_0 n_{ph}(\epsilon_0)d\epsilon_0$. The function $f_{KN}$ results from
the analytical integration of the \cite{Jones1968} Compton kernel, fully taking into 
account Klein-Nishina (KN) effects for an isotropic seed photon field \citep[see][appendix C]{Moder2001},
and $F_{KN}(\gamma)$ represents its convolution with the seed photon field.    
We remark that $F_{KN}$ plays a crucial role in the cooling process,
depending both on the IC regime (Thomson (TH) limit for $4\gamma\epsilon_0<<1$, KN limit for 
$4\gamma\epsilon_0>>1$), and on $\epsilon_0 n_{ph}(\epsilon_0)\propto B^2/R^2$.   


\begin{figure*}[!t]
\centering
\begin{tabular}{l}
\includegraphics[width=18cm,angle=-0]{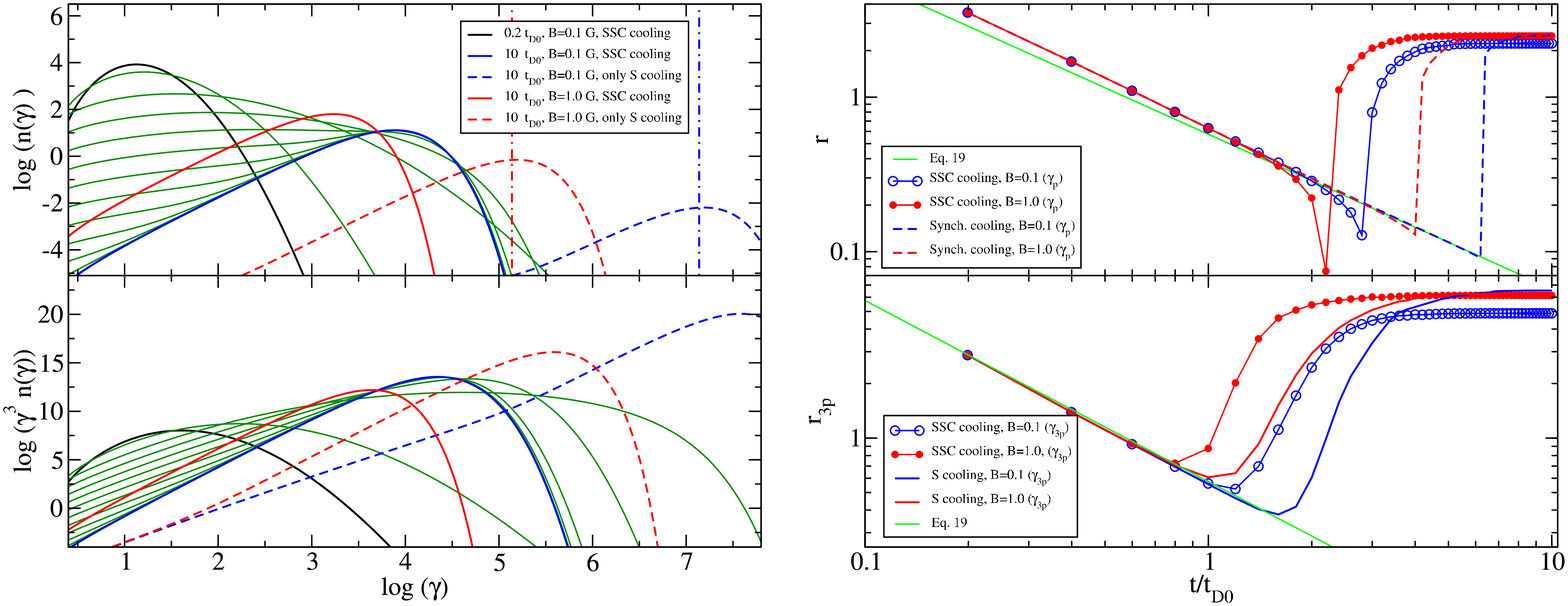}\\

\end{tabular}
\caption{ 
\textit{Left panels: } 
The same as in Fig. \ref{fig:n_evola}, for the case of $R=5\times10^{13}$ cm.
\textit{Right panels:} 
Upper panel: the same as in Fig. \ref{fig:n_evola}, for the case of $R=5\times10^{13}$ cm.
Lower panel: the evolution of the curvature  $r_{3p}$, for the case  $R=5\times10^{13}$ cm.
}
\label{fig:n_evolb}
\end{figure*}

\begin{figure*}[!t]
\centering
\begin{tabular}{l}
\includegraphics[width=18cm,angle=-0]{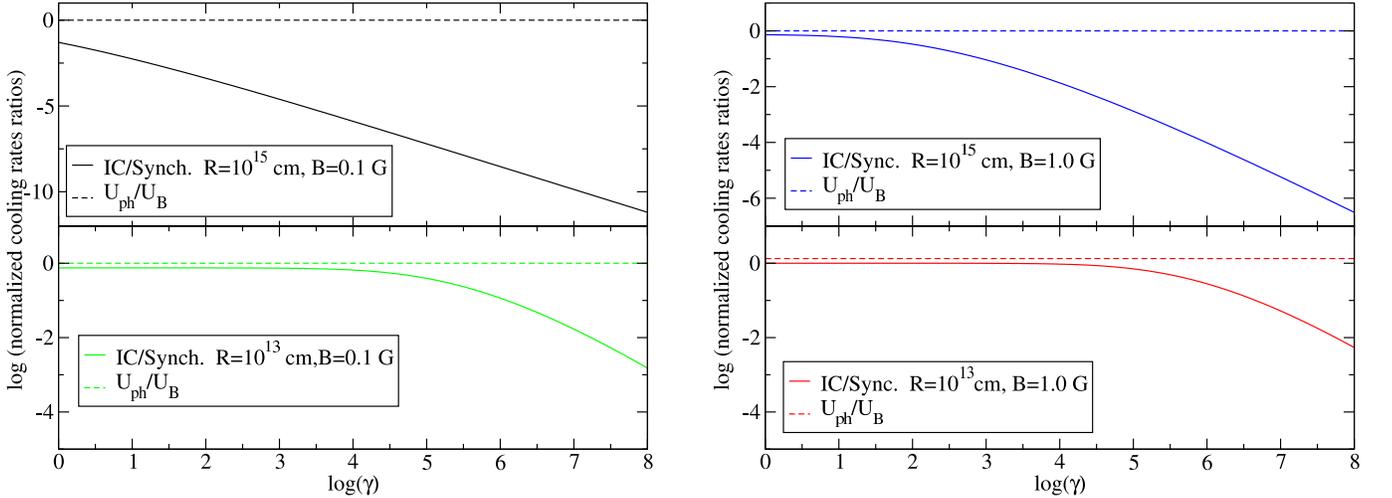}\\

\end{tabular}
\caption{ 
Normalized ratios of electron cooling rates $\dot\gamma_{IC}/\dot\gamma_{Synch.}$ 
(solid lines), and $U_{ph}/U_B$ (dashed lines), as a function of $\gamma$,
for different values of $R$ and $B$, for the case of $q=2$, at the final
step of the evolution.
\textit{Left panels:} top, case of $R=10^{15}$ cm and $B=0.1$ G. Bottom, case of $R=10^{13}$  cm and $B=0.1$ G  
\textit{Right panels:} top, case of $R=10^{15}$ cm and $B=1.0$ G, Bottom, case of $R=10^{13}$ cm and $B=1.0$ G 
}
\label{fig:cooling_rates}
\end{figure*}

\begin{figure*}[!t]
\centering
\begin{tabular}{l}
\includegraphics[width=18cm,angle=-0]{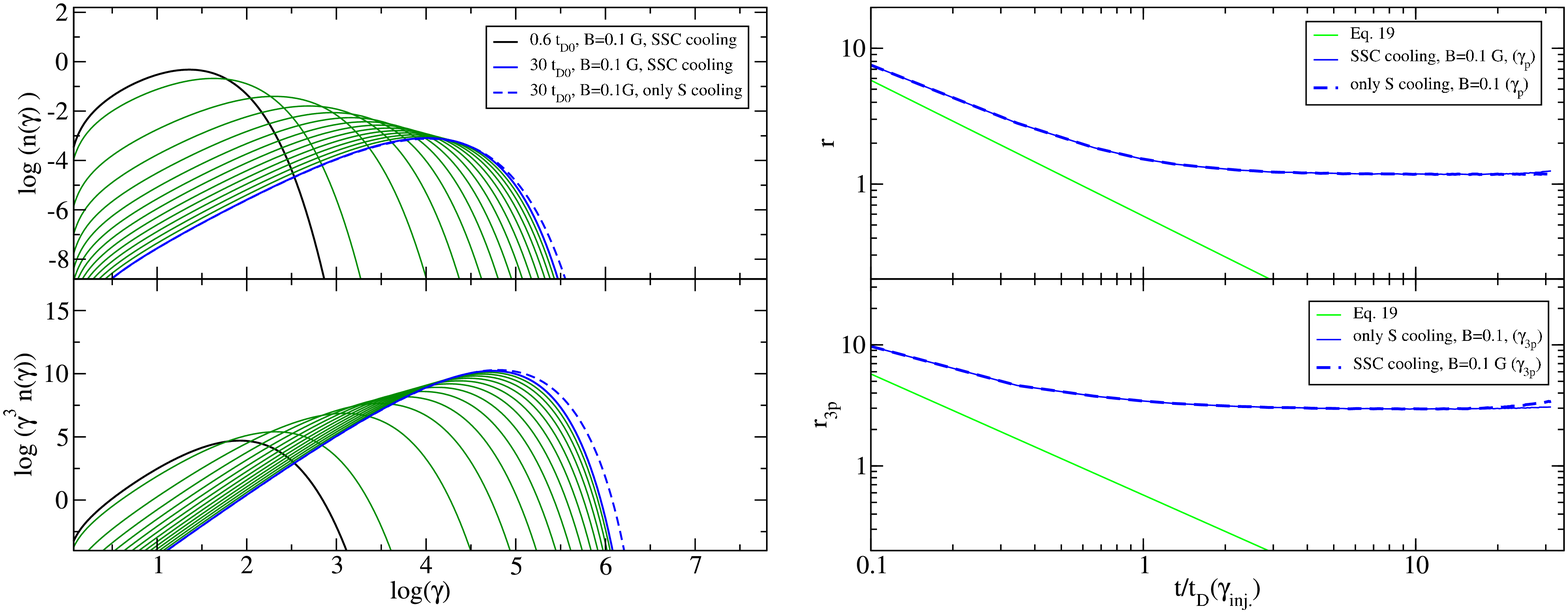}\\

\end{tabular}
\caption{ 
\textit{Left panes: } 
evolution of the particle spectrum with impulsive injection and no escape 
for the case of $R=1\times 10^{15}$ cm, $B=1.0$ and $q=3/2$. 
Since $t_D$ is energy dependent, on the x-axis we plot the ratio $t/ t_{D}(\gamma_{inj})$, where
$t_{D}(\gamma_{inj})$ is the diffusive acceleration time evaluated at the injection
energy $\gamma_{inj}$.
Green solid lines represent the temporal evolution, for $B=0.1$ G, with step of $2.4\times t_{D}(\gamma_0)$.
\textit{Right panels:} 
Evolution of the curvature $r$ (upper) and $r_{3p}$ (lower).
}
\label{fig:n_evolq1.5}
\end{figure*}

Since analytical solutions are possible only for a limited number of cases, 
to follow the complex dependence of the IC cooling term on $n_{ph}(\epsilon_0)$
in a self-consistent way, we must solve the diffusion equation numerically. 
For this purpose we further developed the
numerical code \citep{Tramacere2009,TramacerePhD2007} used to compute numerically
the synchrotron and IC emission, and introduced  it into the numerical solution of the diffusion
equation. In the numerical calculations, we adopted the 
method proposed by Chang \& Cooper(1970) and used the numerical {\it recipe} 
given by \cite{Park1996}. This is a finite difference scheme based on the 
centered difference of the diffusive term, employing weighted differences for the 
advective term.  
We use a $5000$ points energy grid over the range $1.0 \leqslant \gamma
\leqslant10^9$, and a time grid is finely tuned to have a temporal mesh several orders 
of magnitude smaller than typical cooling and acceleration time scales.  
The results from our code were compared, when possible, with known analytical solutions, 
and always found  good agreement.


\subsection{Physical set-up: the relations between $D_p$, and $t_D$ with $\gamma_{max}$  and $R$ }
\label{sec:physical_set_up}
We study the evolution of $n(\gamma)$ and of the curvature term in an
homogeneous spherical geometry, with radius $R$ and an entangled coherent
magnetic field $B$ and a turbulent component $\delta B$, in the two cases
of impulsive and continuous injection with a quasi mono-energetic source
function $Q(\gamma_{inj},t)$ normalised to have a fixed energy input rate:
\begin{equation} L_{inj}=\frac{4}{3}\pi R^3\int\gamma_{inj} m_ec^2
Q(\gamma_{inj},t)d\gamma_{inj} ~~~ (erg/s) 
\end{equation} 
In our approach we don't distinguish the acceleration region from the radiative
one, and during the acceleration process we take into account both synchrotron
and IC cooling. 
According to Eq. \ref{eq:diff_coeff_0}, to determine the order of magnitude of $D_p$
we assume $1>>\delta B/B \simeq 0.1-0.01$ and require Alfven waves to be at least 
mildly relativistic, with $\beta_A\simeq 0.1-0.5$, and their maximum  wavelength
to be much smaller than the accelerator size ($\lambda_{max}<R$). 
To study the effect of IC cooling on the evolution of $n(\gamma)$, we consider two different sizes of the
acceleration region, a compact one ($R=5\times 10^{13}$ cm), and a larger one
($R=1\times 10^{15}$ cm). With this choice of accelerator size we set
$\lambda_{max}\approx 10^{12}$ cm. We stress that the choice of $\lambda_{max}$
constrains the accelerative upper limit through   $\rho_{g}<\lambda_{max}$ leading to
$\gamma_{max}<(\lambda_{max}qB)/m_ec^2$, since particles with larger $\rho_g$ (hence
larger $\gamma$) can't resonate with shorter wavelengths. Taking into account a
coherent magnetic field of the order of $0.1$ G, and $\lambda_{max}\approx
10^{12}$ cm we found that the purely-accelerative efficiency limits the particle
energy to $\gamma_{max}\lesssim 10^{7.5}$. In the left panel of Fig. \ref{fig:t_D} we
plot $t_D$, given by Eq. \ref{eq:diff_coeff}, as a function of $\lambda_{max}$ , 
for the case of $q=2$, $\delta B/B$=0.1, and $\beta_A$=0.5. In this case the
acceleration time is energy independent, and for $\lambda_{max}\approx
10^{12}$ cm it  will be of the order of $t_D=1/D_{p0}\approx 10^4$ s. 
In the case of $q\neq 2$, the
acceleration will have an energy dependence given by Eq. \ref{eq:diff_coeff},
as shown in the right panel of Fig. \ref{fig:t_D} for the case of $q=3/2$.
In this section, we focus on the evolution of the curvature as a function of the
momentum-diffusion term, and therefore  use only the accelerative contributions 
coming from the diffusion terms ($D_p(\gamma), D_A(\gamma)$), neglecting the systematic
extra term $A(\gamma)$.
All the parameters and their numerical values are given in Tab. 1.

\begin{figure*}[!t]
\centering
\begin{tabular}{l}
\includegraphics[width=16cm,angle=-0]{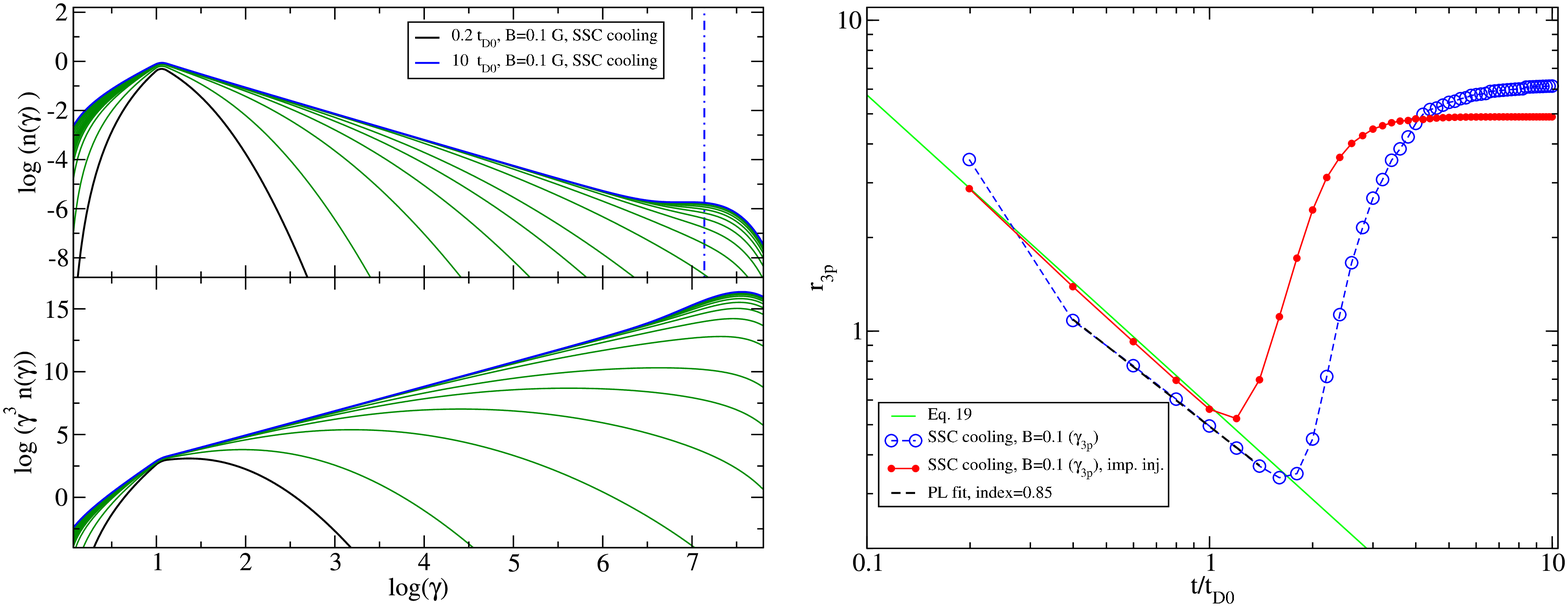}\\
\end{tabular}
\caption{ 
\textit{Left panel: } 
Evolution of the particle spectrum for continuous injection, $R=1\times10^{15}$ 
cm, $B=1.0$ G, and $q=2$.
\textit{Lower panel:}
Evolution of the curvature $r_{3p}$.}
\label{fig:n_evol_cont}
\end{figure*}

\setcounter{table}{0}
\begin{table*}[ht]
\begin{center}
\caption{Parameters values adopted in the numerical solutions of the diffusion equation
for the cases studied in Sec. 4}
\begin{tabular}{ll|l|l|l|l}
\hline
&& \multicolumn{2}{|c}{impulsive inj.}&\multicolumn{2}{|c}{     cont. inj.}\\
\hline
\hline
$R$             &(cm)         &$5\times10^{13}, 1\times10^{15}$&-&- &-   \\
$B$             &(G)          &0.1, 1.0         &-               &-             &-   \\
$L_{inj}$       & (erg/s)     &$10^{39}$       &  -                &$10^{37}$        &-\\
$q$             &             &2               &  3/2             &2       &3/2\\
$t_{D_0}=1/D_{P0}$      & ($s$)       &$1\times 10^4$  &$1\times 10^3$     &$1\times 10^4$   &$1\times 10^3$ \\
$T_{inj}$       & ($s$)     &100           &  -               &$1\times 10^4$               &  -\\
$T_{esc}$       & ($R/c$)     &$\infty $         &  -                &2              & - \\
Duration        & ($s$)       &$1\times 10^5$            &  -                &-  & -\\
$\gamma_{inj}$  &             &10.0            &  -                &10.0               & -\\
\hline

\end{tabular}

\end{center}
\label{tab:sim_par}
\end{table*}

\subsection{Impulsive injection}              
\label{sec:imp_inj}
In the left panels of Fig. \ref{fig:n_evola}, and Fig.  \ref{fig:n_evolb} we plot
the evolution of energy distribution $n(\gamma, t)$ (upper panels)
and of $\gamma^3 n(\gamma, t)$ (lower panels) in the case of the impulsive injection 
without escape, for $q=2$, and for two values of $R$: $1\times
10^{15}$ cm (Fig.  \ref{fig:n_evola}) and $5\times 10^{13}$ cm (Fig. \ref{fig:n_evolb}).
We inject a quasi-monoenergetic electron distribution with $\gamma_{inj}\approx 10$.
The $\gamma^3 n(\gamma, t)$ representation is useful to compare the results concerning
$n(\gamma)$ presented in this section, with those regarding the synchrotron emission
presented in Sec. \ref{sec:ssc_evolution}. We denote by $\gamma_p$ the peak energy 
of $n(\gamma)$ and by $r$ the curvature evaluated by means of a log-parabolic best 
fit over a one decade-broad interval  centered at $\gamma_p$. $\gamma_{3p}$ 
and $r_{3p}$ represent the peak of $\gamma^3 n(\gamma)$ and its curvature, respectively.
In the right panels of Fig. \ref{fig:n_evola}, and \ref{fig:n_evolb}, we report on the corresponding temporal 
evolutions  of the curvatures  
under the effect of both momentum-diffusion and cooling terms.
The solid black line corresponds to $t=0.2\times t_{acc}$, where $t_{acc}=t_{D_0}$ is the
acceleration time due to momentum-diffusion.
As the time increases, the diffusion term acts on the distribution by means
of both $D_{A}$ and $D_{p}$. The effect of the latter is to make the 
distribution broader.

One can distinguish three phases: in
the first one the energy of particles increases
and the curvature parameter decreases following a law $r\propto t^{-1}$ in agreement with 
the statistical scenario of Sec. \ref{sec:Logpar} and with the Eq. \ref{eq:analyt_sol_curv},
independent of the magnetic field strength ($B=1.0$ G and $B=0.1$ G) and of the source size,
because the accelerative contribution dominates over the radiative losses; 
in the second phase, the
radiation losses become relevant and the distribution approaches  the equilibrium with 
an increase of the curvature; in the third phase, 
the balance between acceleration and radiation losses is established and the curvature 
reaches a stable value. 

The equilibrium  distribution reached through  stochastic acceleration, is described by 
a relativistic Maxwellian \citep{Staw2008},
\begin{equation}
n(\gamma)\propto \gamma^2 \exp{\Big[\frac{-1}{f(q,\dot\gamma)} \Big(\frac{\gamma}{\gamma_{eq}}\Big)^{f(q,\dot\gamma)}\Big]},
\label{eq:n_equil}
\end{equation}
where $f(q,\dot\gamma)$ is a function depending on the exponent of the diffusion
coefficient and on the cooling process, and  $\gamma_{eq}$ is the Lorentz factor
that satisfies the condition $t_{cool}(\gamma)=t_{acc}(\gamma)$ and is given by
\begin{equation}
\gamma_{eq}=\frac{1}{t_{acc} C_0(U_B+F_{KN}(\gamma))}~,
\label{eq:t_cool}
\end{equation} 
with $t_{acc}$ equal to the fastest acceleration time scale among $t_A,t_D,t_{DA}$. 
In the case of Compton dominated cooling we have $\gamma_{eq}\propto \frac{R^2}{t_{acc}B^2 f_{KN}}$, whilst in the
case of strong KN regime, or in general for synchrotron dominated cooling, we have 
$\gamma_{eq}\propto \frac{1}{t_{acc}B^2}$.
Using a power-law form for the acceleration terms, and in the case of only synchrotron   
losses (or  any cooling process that can be expressed as a power-law function of 
$\gamma$), it is possible to give an analytic expression of $f(q,\dot\gamma)$ 
\citep{Katar2006, Staw2008}.
The expectation for synchrotron and IC/TH cooling process, and for $q=2$ 
is $f(q,\dot\gamma)=3-q=1$. The curvature resulting from a log-parabolic fit
over a decade centered on $\gamma_p$ is $r\approx 2.5$, and $r_{3p}\approx 6.0$ in
the case of $\gamma_{3p}$

We first discuss the case of $R=10^{15}$ cm (Fig. \ref{fig:n_evola}),
with only synchrotron cooling (dashed lines, left panels).
In terms of behaviour, we note that for the larger value of $B$  (1.0 G, red lines, right panels),
the $r$-$t$ trend departs from the purely accelerative one ($r\propto t^{-1}$, green line right panels)
early (relative to the $B=0.1$ G case, blue lines in the right panels). 
This happens because the synchrotron equilibrium energy (vertical dot-dashed lines, left panels) 
is lower in the 
case of $B=1.0$ G. For both values of $B$, the final values of $r$ are close to the synchrotron 
equilibrium value of $\approx$ 2.5.
When IC cooling is also  taken into account, the final values of the curvature
in $n(\gamma)$ are $r\approx 2.5$ 
and $r\approx 0.6$, for $B=0.1$ G and $B=1.0$, G respectively. 
This difference is due to the different IC cooling regimes for the two cases. 
To show clearly the complexity of the transition from the TH to the KN regime,
and its dependence on $R$ and $B$, in Fig. \ref{fig:cooling_rates} we plot the 
ratio $\dot\gamma_{IC}/\dot\gamma_{Synch.}$ 
(solid lines), and $U_{ph}/U_B$ (dashed lines), as a function of $\gamma$ and normalized
to unity, for the case of $q=2$, for the final step of the temporal evolution.
As long as the ratio $U_{ph}/U_B$ is close to  $\dot\gamma_{IC}/\dot\gamma_{Synch.}$, 
electrons cool in the full  TH regime, and 
$C(\gamma)=C_0\gamma^2 ( U_B+ U_{ph})$.
On the contrary, when the  electrons radiate in the full KN regime $\dot\gamma_{IC}/\dot\gamma_{Synch.}<<U_{ph}/U_B$.
In this case, due to the inefficient KN cooling regime we have $\dot\gamma_{Synch.}>>\dot\gamma_{IC}$,
and the cooling term is dominated by the synchrotron component: $C(\gamma)\approx C_0\gamma^2 U_B$.
In the intermediate cases, it's difficult to estimate analytically the ratio $\dot\gamma_{IC}/\dot\gamma_{Synch.}$.

For $B=1.0$ G, the SSC equilibrium is reached at $\gamma\approx 3\times 10^4$ and the  
SSC cooling occurs between the KN and TH regimes (see top-right panel in
Fig. \ref{fig:cooling_rates}), hence the value of $f$
is different from unity, as predicted for the case of  full IC/TH or synchrotron cooling.
When $B=0.1$ G, the equilibrium energy is $\gamma\approx 10^7$ and electrons are 
in extreme KN cooling (see top-left panel in
Fig. \ref{fig:cooling_rates}), synchrotron losses are much higher than those due 
to IC scattering, and again $r$ reaches the previous value of $\approx 2.5$.
It is also interesting to note the difference in the trends of $r$-$t$ and $r_{3p}$-$t$.
In the latter case, the trend departs form the purely accelerative regime earlier 
(see Fig.\ref{fig:n_evola}, right-lower panel) since the electrons with 
energies close to $\gamma_{3p}$ are more energetic than those close to $\gamma_p$,
and thus have much shorter cooling times.

\begin{figure*}
\centering
\begin{tabular}{l}
\includegraphics[width=18cm,angle=-0]{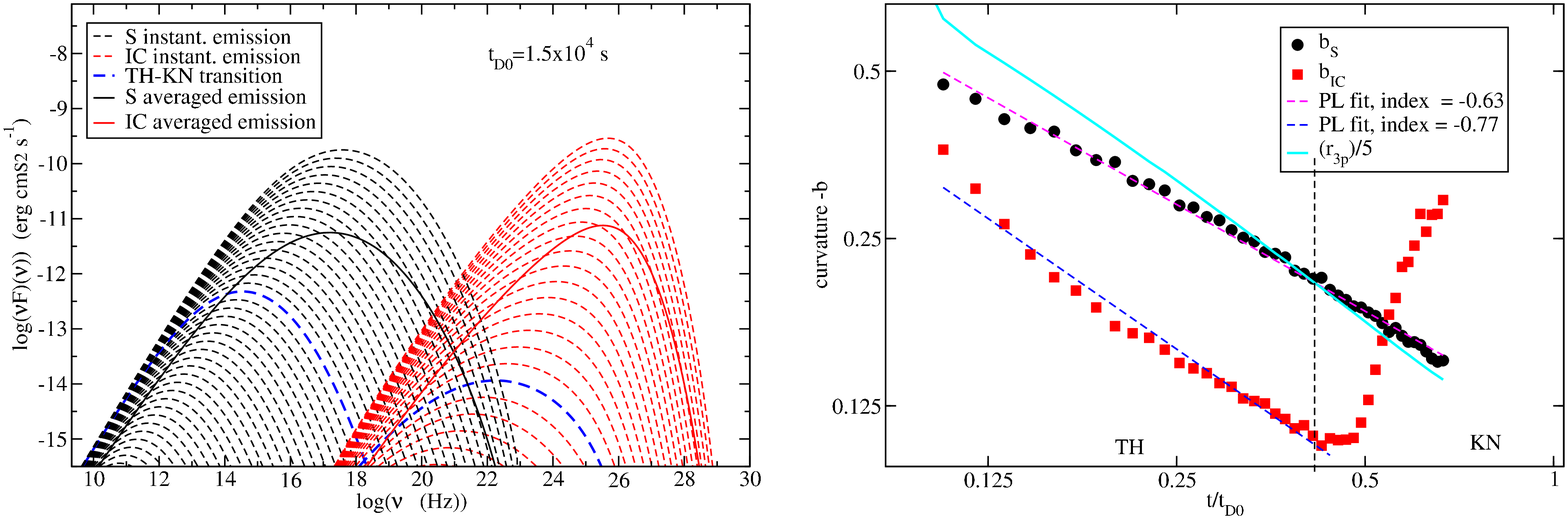}\\
\includegraphics[width=18cm,angle=-0]{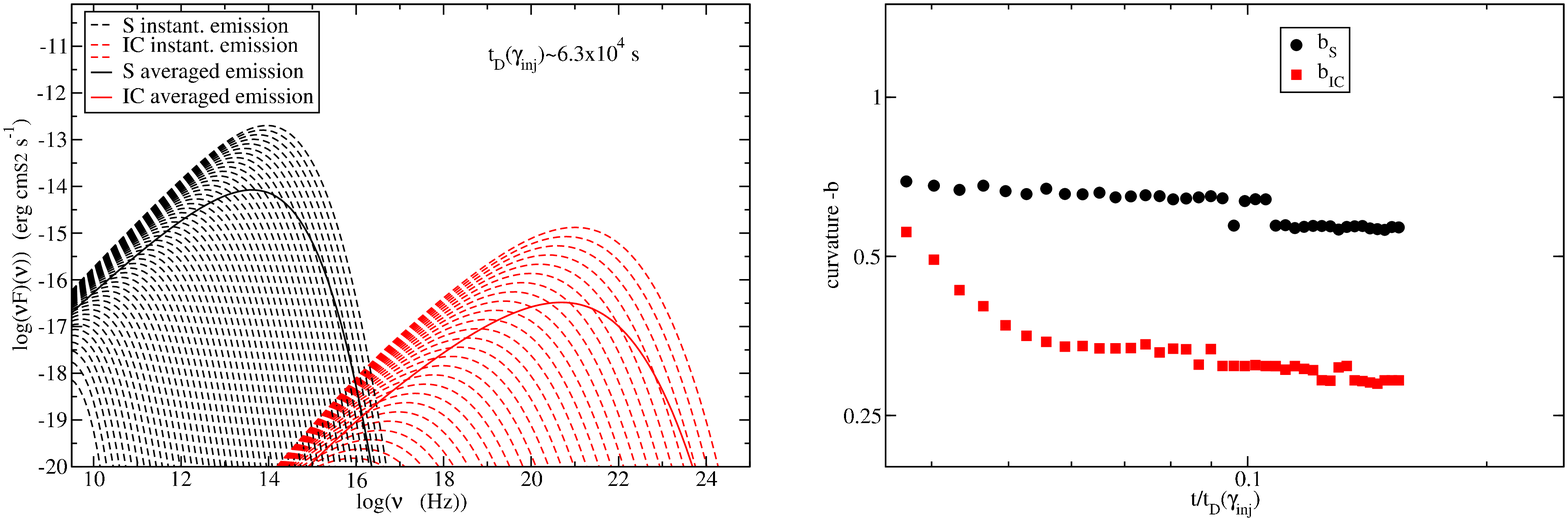}\\

\end{tabular}
\caption{ 
\textit{Left panel:} 
Evolution of synchrotron (black dashed lines) and IC (red 
dashed lines) SEDs, for the case of $t_{D_0}=1.5\times 10^4$ s and $q=2$ (top panel), and
for the case of $t_{D}(\gamma_{inj})\approx 6.3\times 10^4$ s and $q=3/2$ (bottom panel). 
All the other  parameters as reported in Tab. 2. 
The solid lines represent the SEDs averaged
overt the full simulation period, and the blue dashed lines (top panel) represent the SEDs
corresponding to the transition from TH to KN regime.
\textit{Right panel:}
The temporal evolution of $b_s$ (black squares) and $b_c$ (red squares) as a function
of $t/t_{D_0}$, for the case of $q=2$ (top panel), and $q=3/2$ (bottom panels)
The cyan line (top panel) represents the $b_s$ trend predicted for the synchrotron emission
in case of $\delta-$approximation. The dashed lines (top panel) represent the PL best fit
of both  $b_s$ (purple) and $b_c$ (blue) trends.
}
\label{fig:temp_ev_b}
\end{figure*}

The results for the  compact region ($R=5\times 10^{13}$ cm) are plotted
in Fig. \ref{fig:n_evolb}.
Considering that the injected electron luminosity is the same (see Table 1),
we expect a different response from the IC cooling, due to the higher photon
density $n_{ph}(\epsilon_0)$. 
The $r$ evolution for the synchrotron  cooling case is similar to the 
previous case,
while for the SSC emission, both for the case of $B=1.0$ G and $B=0.1$ G, 
the final value of $r$ is about  2.5. 
This is due to the larger photon density which moves the IC scattering into the
TH regime also for the case of $B=0.1$ G (compare bottom-left to top-left panel
in Fig. \ref{fig:cooling_rates}), hence $n(\gamma)$ approaches the solution of Eq. \ref{eq:n_equil}  
in the case of $f=1$.

In Fig. \ref{fig:n_evolq1.5}, we show the temporal evolution for the case
of $q=3/2$ ($R=1.0\times10^{15}$ cm, B=0.1 G). 
In this case,
contrary to the $q=2$ case, the acceleration time $t_D$ is energy dependent,
hence we study the evolution of $r$ as a function of $t/ t_{D}(\gamma_{inj})$, where
$t_{D}(\gamma_{inj})$ is the diffusive acceleration time evaluated at the injection
energy $\gamma_{inj}$.
The energy dependence of $t_{D}$ affects the evolution of $r$, and  the shape  of the equilibrium 
distribution, indeed, the $r$-$t$, and $r_{3p}$-$t$ trends  show  different behaviour compared to
the case of $q=2$.  The equilibrium curvature is reached  for
$t\gtrsim 1\times t_{D}(\gamma_inj)$, and the two  equilibrium curvature values are
$r\approx 1.2$ and $r_{3p}\approx 3.0$, roughly half of those found for the case
of $q=2$, and in agreement with the result from the MC. We note that,   
the curvature obtained by means of a log-parabolic fit of Eq. \ref{eq:n_equil}, 
for the case $q=3/2$ (namely $f=1.5$), is $r\approx 3.7$. Hence, both the MC and 
the numerical solution of the diffusion equation give a result different from that predicted by the
analytical solution in Eq. \ref{eq:n_equil}.
\\
\\

\subsection{Continuous Injection}
\label{sec:cont_inj}
The case of continuous injection (see Fig. \ref{fig:n_evol_cont}) is more complex.
The distribution developes a low-energy power-law tail, but a log-parabolic 
bending, driven by the diffusion, is still present at high energies,
hence we evaluate the curvature only at $\gamma_{3p}$,
(i.e.  the representation useful to compare to the synchrotron SED emission). 
Spectral curvatures are generally milder than the impulsive injection. In 
the left panel
of Fig. \ref{fig:n_evol_cont} we plot the $r$-$t$ trend both for the case
of impulsive (red lines) and continuous (blue lines) injection, the curvature
in the continuous injection case are systematically lower in the pre-equilibrium
phases, and in the acceleration-dominated stage the trend is again 
consistent with the ``hard-sphere'' approximation and  statistical approaches.
The slope of the electron distribution in the power-law tail is $\approx 1.06$, 
in  good agreement with the predicted one $\approx 1+ t_{min-acc}/(2t_{esc}) = 1.075$, 
consistent with the results of \cite{Katar2006}.

\section{Evolution of the spectral parameters of synchrotron and IC emission }
\label{sec:ssc_evolution}
The most relevant parameters describing the SED of SSC sources provided by observations 
are the peak energies and curvatures of the synchrotron and IC components.
We denote these curvature parameters by $b_{s}$ and $b_{c}$, respectively, 
and by 
$E_{s}$, $E_{c}$ and $S_{s}$, $S_{c}$, we denote the corresponding SED peak 
energies, and flux values.
We use $\nu_s$ and $\nu_c$ to indicate the corresponding SED peak frequencies.
In the following, we describe the results of the relations between these
parameters assuming that electrons are injected into the acceleration region 
with a quasi mono-energetic spectrum with $\gamma_{inj}\approx 10$, and using
an injection time of $10^4$ s.   
We use the same working hypothesis for the momentum diffusion coefficient as
in Sec. \ref{sec:physical_set_up}, and add a systematic acceleration time 
for the first order  process $t_A=1.5\times 10^3$~s, in order to produce $E_s$ 
values ranging between  optical and hard X-ray energies. We set the radius of 
the region at $R=2\times10^{15}$ cm and the same duration  
for the injection and acceleration processes, namely $10^4$ s. 
We varied the other parameters of the model, $B$, $q$, and $D_{p0}$ to verify how
they affect the relation between the observable ones.
All the parameters and their variation ranges are summarised in Tab. 2.

\setcounter{table}{1}
\begin{table}[ht]
\begin{center}
\caption{Parameters values adopted in the numerical solutions of the diffusion equation
for the cases studied in Sec. 5}
\begin{tabular}{ll|l}
\hline
\multicolumn{2}{l|}{parameter }&{range}\\
\hline
\hline
$R$             &(cm)         &$2\times10^{15}$\\
$B$             &(G)          &[0.01 - 1.0]              \\
$L_{inj}$       & (erg/s)     &$10^{38}$    \\
$q$             &             &[3/2 - 2]        \\
$t_A$           & (s)       &$1.8\times 10^3$ \\
$t_{D_0}=1/D_{P0}$      & (s)       &$[1.5 - 25]\times 10^4$  \\
$T_{inj}$       & (s)     &$10^4$            \\
$T_{esc}$       & ($R/c$)     &2.0      \\
Duration        & (s)     &$10^4$\\
$\gamma_{inj}$  &             &10.0 \\
\hline
\end{tabular}
\end{center}
\label{tab:sim_spectra}
\end{table}


A phenomenological approach, based on the $\delta$-function approximation
\citep{Tramacere2007Mrk421,Tramacere2009,Massaro2006,Massaro2004}, is useful to
address the expected relation between the curvature parameters and their
connections with the peak energies and flux values.
According to the standard synchrotron theory (e.g. \cite{Ryb1986}), 
in the $\delta-$function approximation, the synchrotron SED peak value and 
the corresponding peak energy can be expressed by the following relations:
\begin{eqnarray}
\label{eq:S_peak_val}
S_s(E_s) &\propto& n(\gamma_{3p})\gamma_{3p}^3 B^2\delta^4\\
E_s    &\propto&   \gamma_{3p}^2 B \delta. \nonumber
\end{eqnarray}
which implies
\begin{equation}
S_s \propto (E_s)^\alpha,
\label{eq:Sp_Ep}
\end{equation}
where $\alpha=1.5$ applies for changes of $\gamma_{3p}$ leaving constant
$n(\gamma_{3p})$, $\alpha=2$ for variations of $B$ only, and $\alpha=4$ when the
main driver is $\delta$. For a log-parabolic shaped $n(\gamma)$ we have:
\begin{equation}
\log(\gamma_{3p})=\log(\gamma_{p}) +\frac{3}{2r}
\label{eq:gamma_{3p}}
\end{equation}
and, using the relation $b_s\approx r/5$ \citep{Massaro2004}, 
or, more precisely,  from the analysis presented in Sec. \ref{sec:imp_inj},
$b_s\simeq r_{3p}/5$.
It follows: 
\begin{equation}
\log(E_s) \propto 2\log(\gamma_{p})+ \frac{3}{5b_s}~~~. 
\end{equation}
The relation between $b_s$ and $E_s$ is:
\begin{equation}
b_s= \frac{a}{\log(E_s/E_0)}
\label{eq:Ep_b}
\end{equation}
with $a=3/5=0.6$

The spectral properties of the IC emission are more complex, depending on the 
transition from the TH to the KN regime
\citep[see][for a detailed discussion]{Massaro2006}. 
In the former case, the curvature is close to that of the synchrotron emission, but 
systematically smaller due to the energy redistribution by the scattering process.
In the transition to the KN regime, the energy of IC photons will approach $\gamma m_ec^2$,
hence the IC spectral shape will reflect that of the high-energy tail of $n(\gamma)$,
and the curvature $b_{c}$ will be closer to that of the electrons. 
Then, provided the IC scattering happens in the TH regime, the trends involving  $b_{c}$  
are expected to be similar to those of $b_{s}$, but showing systematically $b_{c} < b_{s}$. 
As the KN regime is approached, $b_{c}$ changes differently from $b_{s}$, 
converging towards $r$. 

\subsection{Temporal evolution of $b_s$, and $b_{c}$}
We compute the evolution of $b_s$ and $b_{c}$, as a function of the time, for
the case of $t_{D_0}=1.5\times 10^4$ s, $B=0.1$ G, and $q=2$, using temporal
mesh of 2 s.
We plot in the top-left panel of Fig. \ref{fig:temp_ev_b} the instantaneous SEDs 
at steps of 200 s: the solid lines represent the synchrotron and IC SEDs averaged 
over the full duration of the acceleration process ($10^4$ s).
As the time is increased, the peak energy of both the synchrotron and IC SEDs moves towards 
higher energies with a broadening of the spectral distribution.
The corresponding evolution of curvature parameters is reported in the top-right panel:
$b_s$ has a trend similar to that of the electron distribution,  
with $b_s \propto (t/t_{D0})^{-\alpha}$, and $\alpha\simeq0.6$ (
for comparison the cyan solid line represents the $r_{3p}/5$ trend, as 
predicted by the S $\delta-$approximation).
The trend of $b_{c}$, as expected, is more complex because of the transition 
from TH to KN regime.
For $t/t_{acc}\lesssim0.4$, it follows the same trend of $b_s$ but with  
systematically lower values. For $t/t_{acc}\gtrsim 0.4$, when the TH-KN transition 
occurs, $b_{c}$ increases with time, approaching towards the electron
curvature $r$ value. 
This transition starts for values of $E_s\approx 5\times10^{-3}$ keV ($\nu_s\approx 10^{14}$ Hz),
and $E_c\approx 0.05$ GeV  ($\nu_c\approx 10^{22}$ Hz); and the corresponding SEDs are 
plotted by blue thick-dashed lines in the left panel of Fig. \ref{fig:temp_ev_b}.

In the bottom panels of Fig. \ref{fig:temp_ev_b}, we show the case of
$q=3/2$. The synchrotron curvature  quickly approaches  the equilibrium value of
$b_s\approx0.6$, consistent with the equilibrium value $r_{3p}\approx 3.0$ 
discussed in Sec. \ref{sec:imp_inj}. In this case we do not observe the TH/KN
transition in the IC curvature, since the lower values of $E_s$ and $E_c$,
keep the IC scattering mainly in the TH regime.

\begin{figure*}
\centering
\begin{tabular}{l}
\includegraphics[width=18cm,angle=-0]{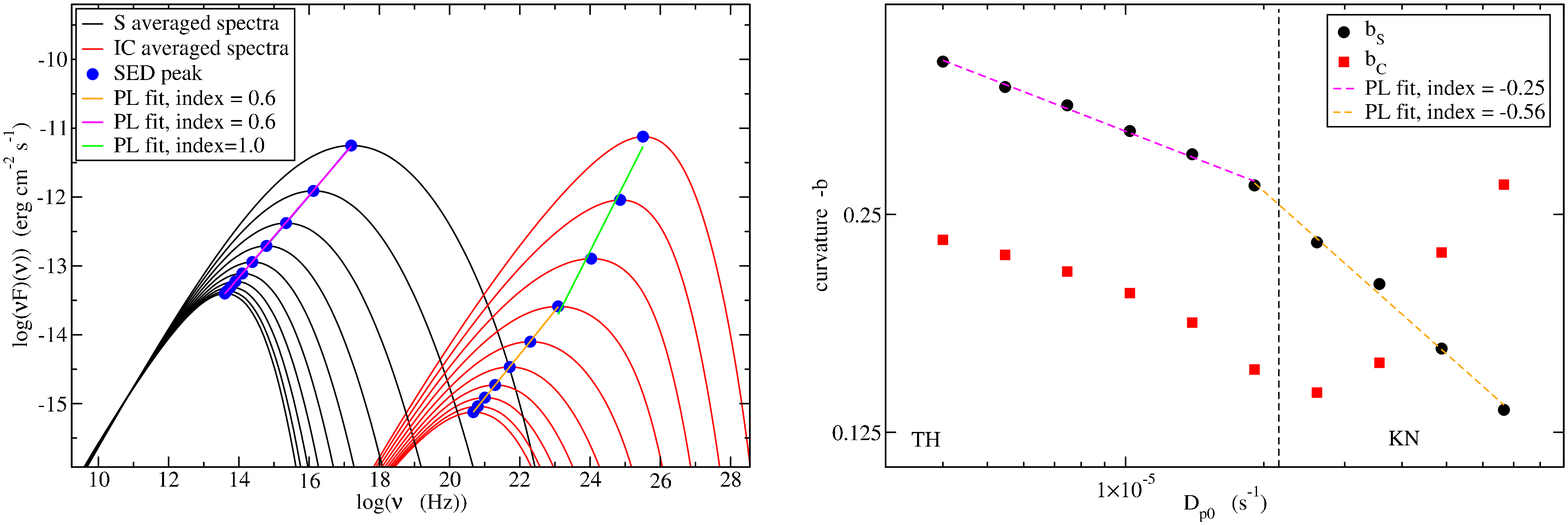}\\
\includegraphics[width=18cm,angle=-0]{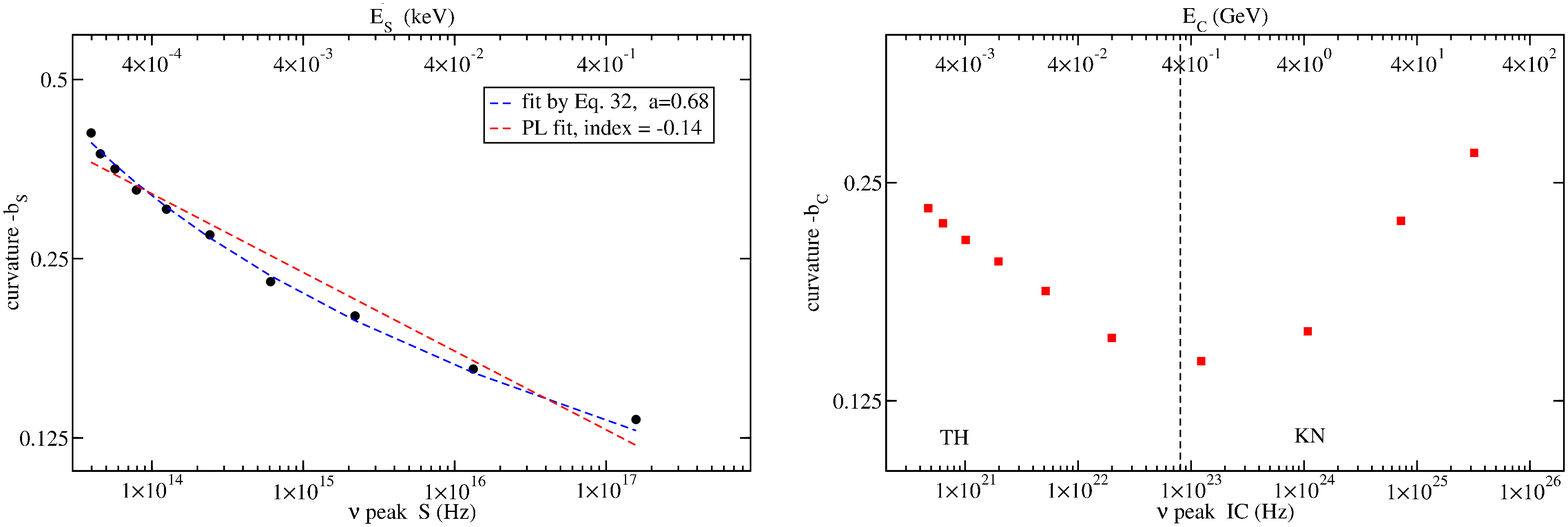}\\
\end{tabular}
\caption{ 
\textit{Upper left panels:} synchrotron (red lines), and IC (red lines) average SEDs
for each  different value of $t_{D_0}$ in the range reported in Tab. 2, with $q=2$. 
Blue points
represent the position of $E_{S,C}$ and $S_{S,C}$. The purple, orange, and green  line
represent the PL best fit of the $E_{S}$-$S_{S}$ and $E_{C}$-$S_{C}$  trends.
\textit{Upper right panel:}
$b_s$ and $b_c$, for each average SED in the right panel, as a function
of $D_{p0}$. Dashed lines represent the PL best fit of the $b$-$D_{p0}$ trend.
\textit{Lower left  panel:} the $b_s$-$E_s$ trend obtained by means of a log-parabolic
best fit of the averaged SEDs plotted in the upper right panel
\textit{Lower left  panel:} same as in the lower left panel, for the $b_c$-$E_c$}

\label{fig:Ep_vs_b_D_trend}
\end{figure*}

\begin{figure*}
\centering
\begin{tabular}{l}
\includegraphics[width=18cm,angle=-0]{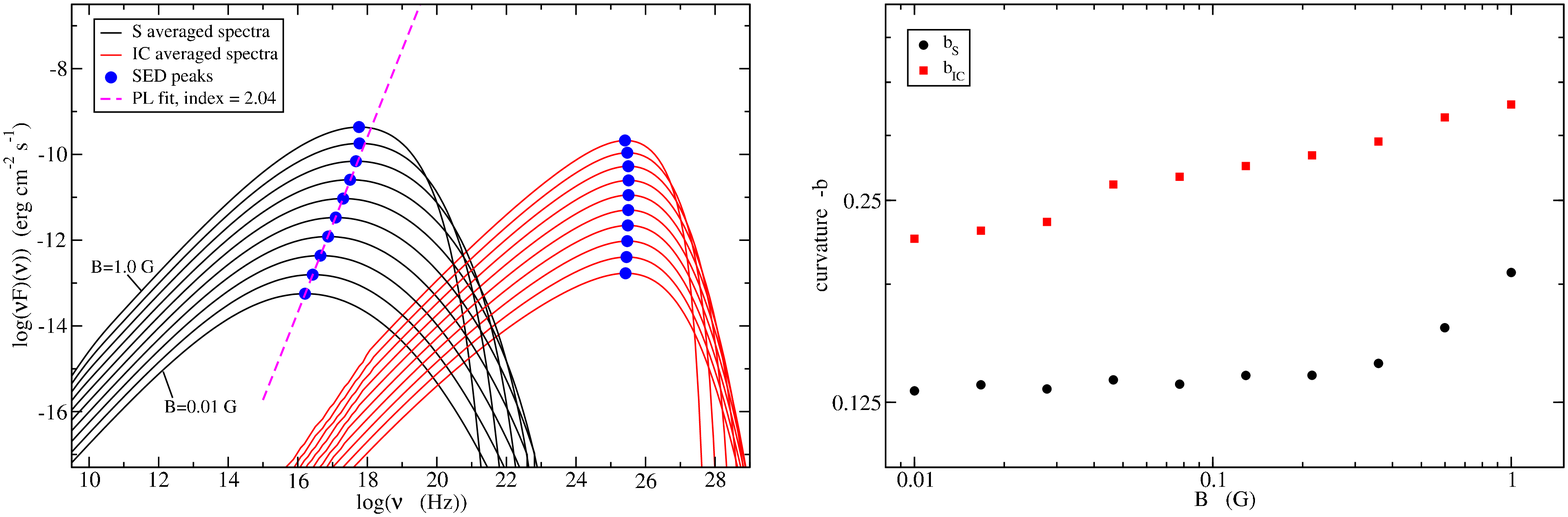}\\
\includegraphics[width=18cm,angle=-0]{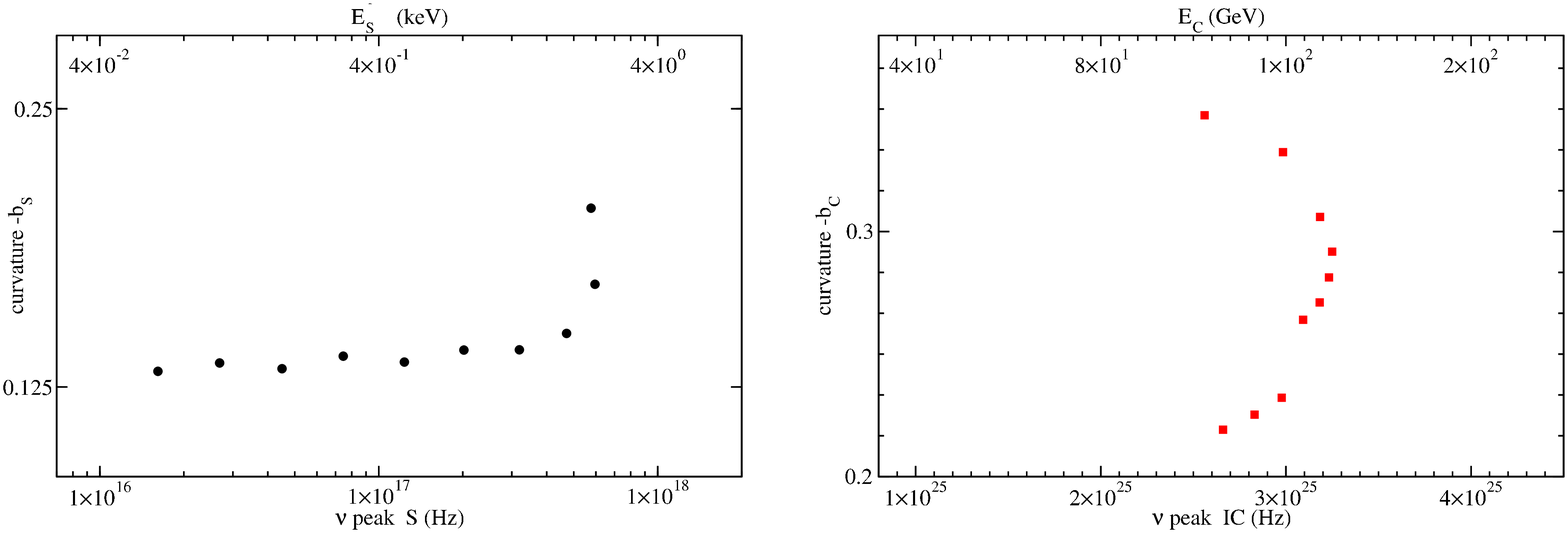}\\
\end{tabular}
\caption{Same as in Fig. \ref{fig:Ep_vs_b_D_trend}, for different values
of $B$ in the range reported in Tab. 2.}
\label{fig:Ep_vs_b_B_trend}
\end{figure*}

\subsection{$E_s$-$S_s$ and $E_s$-$b_s$ as a function of $D_{p0}$ and $q$} 
\label{sec:D-q-spec-trend}
The other parameter affecting the evolution of the spectral distributions is the
diffusion coefficient $D_{p0}$ (see Eq. 15) which we assume to vary in the range 
[$1.5\times 10^4,2.4\times10^5$] s$^{-1}$, studying how the
main spectral parameters change.  In the left top panel of Fig.
\ref{fig:Ep_vs_b_D_trend}, we plot  averaged SEDs for each different value of
$D_{p0}$. The top-right panel shows the trend of $b_{c}$ vs.
$D_{p0}$. As expected, for larger values of $D_{p0}$, the curvature
measured at the peak energy is smaller. The trend is described by a PL with
an exponent of about $-0.6$ for $D_{p0}\lesssim 2\times 10^{-5}$ s$^{-1}$, and
with an exponent of about  $-0.25$ for $D_{p0}\gtrsim 2\times 10^{-5}$ s$^{-1}$.
This break clearly shows  the transition between the TH and KN regime (marked by a vertical dashed
line), indeed it happens for the same values of $D_{p0}$ corresponding to the TH/KN transition  in 
both the $D_{p0}$-$b_c$ trend,  and the $E_c$-$b_c$ plot (occurring  
at $E_c\approx$ 1 GeV, see bottom-right panel in Fig. \ref{fig:Ep_vs_b_D_trend}).
The break in the $D_{p0}$-$b_s$ trend happens when electrons 
radiating at $E_s$ enter the KN cooling 
region, hence, due to the lower cooling level (compared to the TH cooling regime, 
on the left side of the vertical dashed line), the curvature decreases.

Blue filled circles in the top-left panel represent the peak positions for
both SED components. For the synchrotron component, according to Eq.
\ref{eq:Sp_Ep}, the exponent $\alpha$ in the case of $n(\gamma_{3})=$const,
should be 1.5, while the results of the computations give $\alpha=0.6$. This
difference is due to the fact that we inject in the mono-energetic initial
distribution, always the same total power that correspond to the same number 
of particles.
When the peak energy increases the distribution becomes broader, implying that
the same total number of particles is spread over a larger energy interval and
the number of particles contributing to the synchrotron peak emission decreases.
Consequently, the $S_s$-$E_s$ trend gets softer compared to the predicted value
of $1.5$.

We verified quantitatively this effect by computing the trend $n(\gamma_{3p})$ vs
$\gamma_{3p}^2$, and found a power-law relation with an exponent of about 0.98, in 
nice agreement with the difference between the exponent of 1.5 and that
resulting in our simulations. In the bottom panels of Fig.
\ref{fig:Ep_vs_b_D_trend} we plot $b_{s}$ vs $E_s$ (left), and
$b_{c}$ vs $E_c$ (right).  
The $S_c$-$E_c$ relation can be fitted by a power-law (orange line, top-left panel
in Fig. \ref{fig:Ep_vs_b_D_trend}) with the same exponent 
of the $E_s$-$S_s$ relation, as long as the IC scattering, at $E_c$ and above,
happens in TH regime. When the KN suppression becomes relevant (green line, top-left panel
in Fig. \ref{fig:Ep_vs_b_D_trend}), the exponent is larger, and is close to unity.

The synchrotron trend (bottom-left panel in Fig. \ref{fig:Ep_vs_b_D_trend}) clearly  shows the expected
anti-correlation between the peak energy and the spectral curvature, which is well
fit by the function given in Eq. \ref{eq:Ep_b}, with $a=0.68$, not very
different from 0.6, obtained for the $\delta$-function approximation of the synchrotron
emission, and assuming that $n(\gamma)$ has a purely log-parabolic shape. A simple 
power-law fit of the same points returns an exponent $-0.14$. 

We also investigate the role of $q$ on the spectral evolution, 
setting its variation range to [$3/2-2$], i.e. 
from the Kraichnan to the ``hard-sphere'' case.
The relations between the spectral parameters are very similar to those 
found in the previous case with $S_{s} \propto E_s^{0.6}$, and 
$S_c \propto E_c^{0.9}$.
Also in this case, the synchrotron component follows the expectation with a lower 
curvature for harder turbulence spectra, and the IC trend shows the 
transition from TH to KN regime. 
The power-law best fit of $S(E_{s})$ vs $E_s$ gives $a=0.88$, larger than that 
obtained for the case of $D_{p0}$. 
In fact, for values of $q$ lower than 2, corresponding to less turbulence 
and hence diffusion, the curvature gets higher values and the peak energy 
lower values, compared to the ``hard-sphere" case. 
The power-law fit for $b_s$ vs $E_s$ returns an exponent of $-0.16$, practically
coincident with the previous one, indicating that the average properties of these
parameters are the same in both the $q$ and $D_{p0}$ cases.

\subsection{$E_{s,c}$-$S_{s,c}$ and $E_{s,c}$-$b_{s,c}$ as a function of $B$}
\label{sec:B-spec-trend}
The magnetic field $B$ drives the radiative losses, which affect the
evolution of the spectral parameters. 
In Sec. \ref{sec:numeric}, we showed that different cooling conditions, and the transition 
from TH to KN,  can determine very different values 
of $\gamma_{eq}$, for the same acceleration conditions. 
Assuming  that the acceleration time scale  is independent 
of the magnetic field, Eq. \ref{eq:t_cool} shows that  $\gamma_{eq}\propto1/B^2$, implying that, as long 
as $B$  is small enough   to result in $\gamma_{3p}<<\gamma_{eq}$, the evolution of 
$n(\gamma)$ around the peak value is dominated by the acceleration terms, while, for 
values of $B$ resulting in $\gamma_{3p}\gtrsim\gamma_{eq}$ the evolution obtains a 
notable contribution due to cooling.
In the top-left panel of Fig. \ref{fig:Ep_vs_b_B_trend},
we plot the averaged SEDs. 
According to Eq. \ref{eq:S_peak_val}, and \ref{eq:Sp_Ep}, the synchrotron peak value should scale as  
$S_s \propto (E_s)^2$. 
Indeed, for values of $B\lesssim 0.2$ G we obtain an exponent equal to 2.04, very close 
to the value found with the $\delta-$ approximation. 
For higher values of the magnetic field, $E_s$ is anti-correlated with $S_s$. 
This behaviour represents a cooling signature due to the decreasing value $\gamma_{eq}$
for increasing $B$ values,  with $\gamma_{eq}$ getting closer to $\gamma_{3p}$.
This is confirmed both by the shape of the synchrotron SEDs and by the $b_s$-$B$ plot
(top-right panel in Fig. \ref{fig:Ep_vs_b_B_trend}). 
Indeed, S SEDs  for $B\gtrsim 0.2$ G, exhibit an exponential decay, meaning that the 
distributions have reached, or are close to reaching,  the equilibrium energy. 
Consistently with the S shape evolution, the $b_s$-$B$ relation shows an almost stable value of $b_s$ for $B\lesssim 0.2$ G,
and an increasing trend for $B\gtrsim 0.2$ G. This change, in both the $S_s$-$E_s$ and $b_s$-$B$ trends, is interesting and can provide a useful phenomenological 
tool to understand the evolution of non-thermal sources.
Another interesting feature is shown in the $S_c$-$E_c$ plot: for $B\lesssim 0.2$ G the IC peak
energy is practically constant, as expected in the KN limit from the kinematical limit
relating the scattered photons energy to that of the electrons:
$h\nu_{IC}\approx \gamma m_ec^2$. 
In fact, photons at energies $\approx E_c$  are produced in the KN regime, and for $B\lesssim 0.2$ G the
electron peak energy $\gamma_{3p}$ is constant, so $E_c$ must also be constant. 
For $B\gtrsim 0.2$ G, $\gamma_{3p}$ decreases because of cooling, and, accordingly, 
$E_c$ also  decreases. 
This is another interesting test that can provide a probe for $B$ driven flares evolving to
the KN regime.
The $E_s$-$b_s$ plot in the bottom-left panel of Fig  \ref{fig:Ep_vs_b_B_trend},
confirms the cooling signature discussed above, showing $b_s$ uncorrelated
with $E_s$ as long as $\gamma_{3p}<<\gamma_{eq}$, and an increasing value
of $b_s$ with $E_s$ almost stable, when $\gamma_{3p}\gtrsim\gamma_{eq}$.


\section{Spectral evolution of high energy flares of bright HBL objects}
\label{sec:fit_data}
The previous considerations on the spectral evolution of SSC sources, in which
high energy electrons are accelerated in a relatively short timescale by 
stochastic processes, can be successfully applied to describe the behaviour of 
some bright HBLs objects. 
These sources are, in fact, characterised by having the
synchrotron peak in the UV/X-ray range and the IC peak in  $\gamma$ rays up to
TeV energies.  Several flares, observed simultaneously in both these ranges,
exhibited SEDs very well described by a log-parabolic law, whose parameters, 
particularly their curvature, are estimated with  high accuracy. A similar
analysis for low-energy peaked BL Lac objects is much more difficult because the
peak of their synchrotron component is typically in the infrared range and the available
simultaneous multifrequency data are extremely few.
\citet{Tramacere2007Mrk421,Tramacere2009} and \cite{TramacerePhD2007} pointed
out that the observed anticorrelation between $E_s$ and $b_s$  in the synchrotron
SED of Mkr 421, can provide a clear signature of a stochastic component in the
acceleration process. In the same analysis, these authors presented also an interesting correlation 
between $E_s$ and $S_s$.
\cite{Massaro2008} found that the $E_s$-$b_s$ and $E_s$-$S_s$ trends hold also for a
larger sample of eleven HBLs, making stronger the hypothesis that a common accelerative mechanism may drive 
such physical processes for this class of active galactic nuclei . To give a theoretical
framework to these phenomenological relations, we try to reproduce both the $E_s$-$b_s$ 
and $E_s$-$S_s$ relations derived from the data of the aforementioned papers. 
In the following we will consider the data of Mrk 421 from 
\citet{Tramacere2007Mrk421,Tramacere2009}, collected over a period of 13 years, 
and of six HBL objects from \cite{Massaro2008}: 
Mrk 180, Mrk 501, PKS 0548-322, PKS 1959-650, 1H 1426+428, 
covering a period of about 11 years and including both quiescent and flaring states.
The  sources from  \cite{Massaro2008} were chosen because the data are good 
enough to safely constrain both curvature and $E_s$ values, and because the 
observed variations of the sample luminosity are compatible with the assumption 
to be driven by changes of $E_s$.

Following  the analysis presented in Sec. \ref{sec:ssc_evolution}, we consider two 
scenarios in which these trends are driven by the momentum-diffusion term. 
In the first case, the momentum diffusion changes because of variations of $D_{p0}$,
due to changes of $\delta B/B$ or $\beta_A$,
but the turbulence spectrum ($q=2$) remains stable.  
In the second scenario the turbulence spectrum is variable with $q$ ranging in
[3/2, 2]. 
We use the same method described in Sect. \ref{sec:ssc_evolution}, to compute the averaged
SEDs for each value of $D_p$ (or $q$); computations are performed for three values of the 
magnetic field B = 0.05, 0.1, and 0.2 G.
All the model parameters are summarised in Tab. 3.

The comparison with the data can be affected by an observational bias due to the limited
energy range of detectors.
In fact, when the peak energy is close to the limits the curvature is not well estimated
because one can use only a portion of the parabola below or above the peak.
Generally, curvatures lower than the actual ones are obtained. 
The energy range [0.5, 100.0] keV is the typical spectral window covered by X-ray and
hard-X-ray detectors. 
In our analysis we used this fixed window to take into account this possible bias in the 
observed data when $E_s$ is variable.

\begin{figure*}
\centering
\begin{tabular}{ll}
\includegraphics[width=9cm,angle=-0]{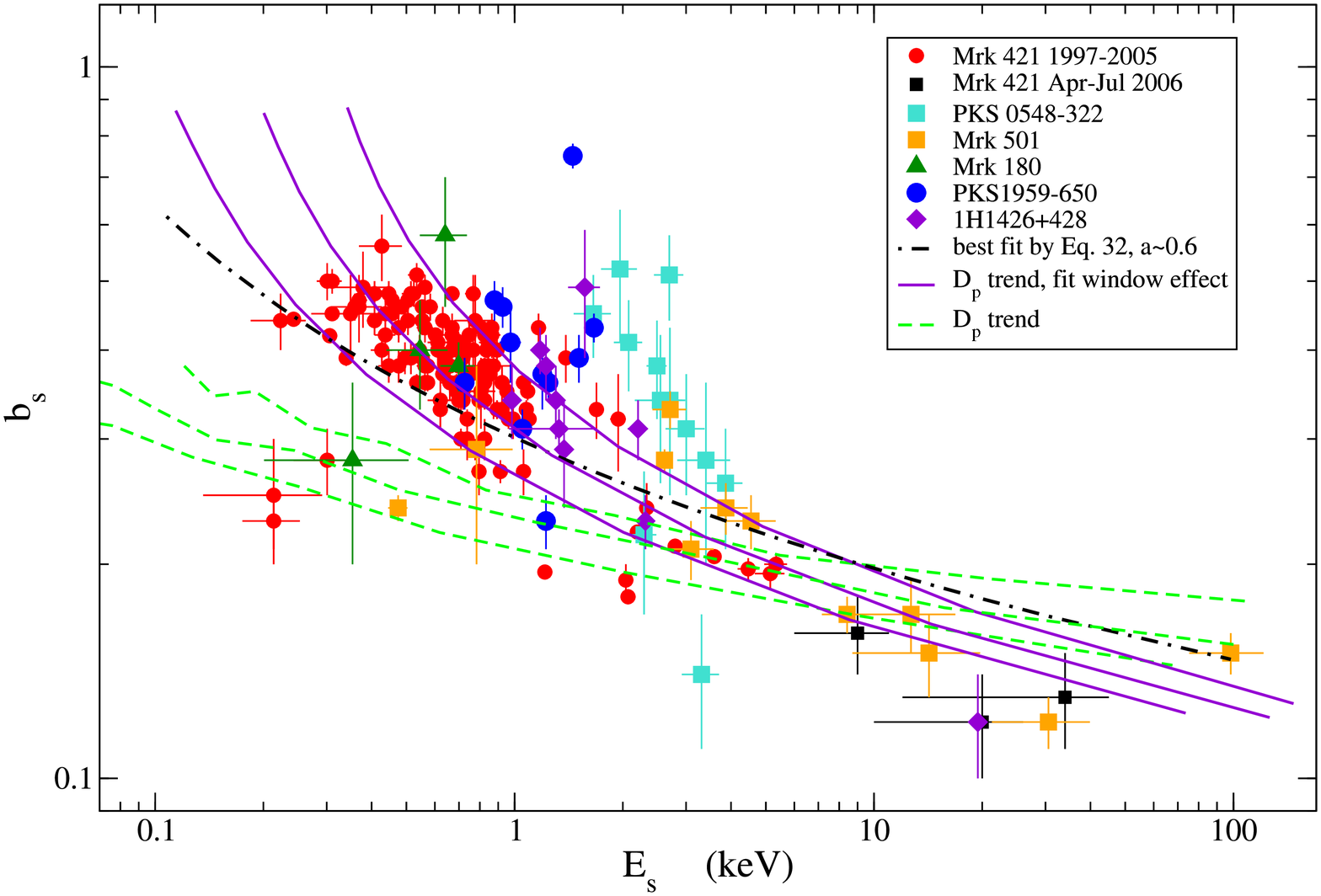}&
\includegraphics[width=9cm,angle=-0]{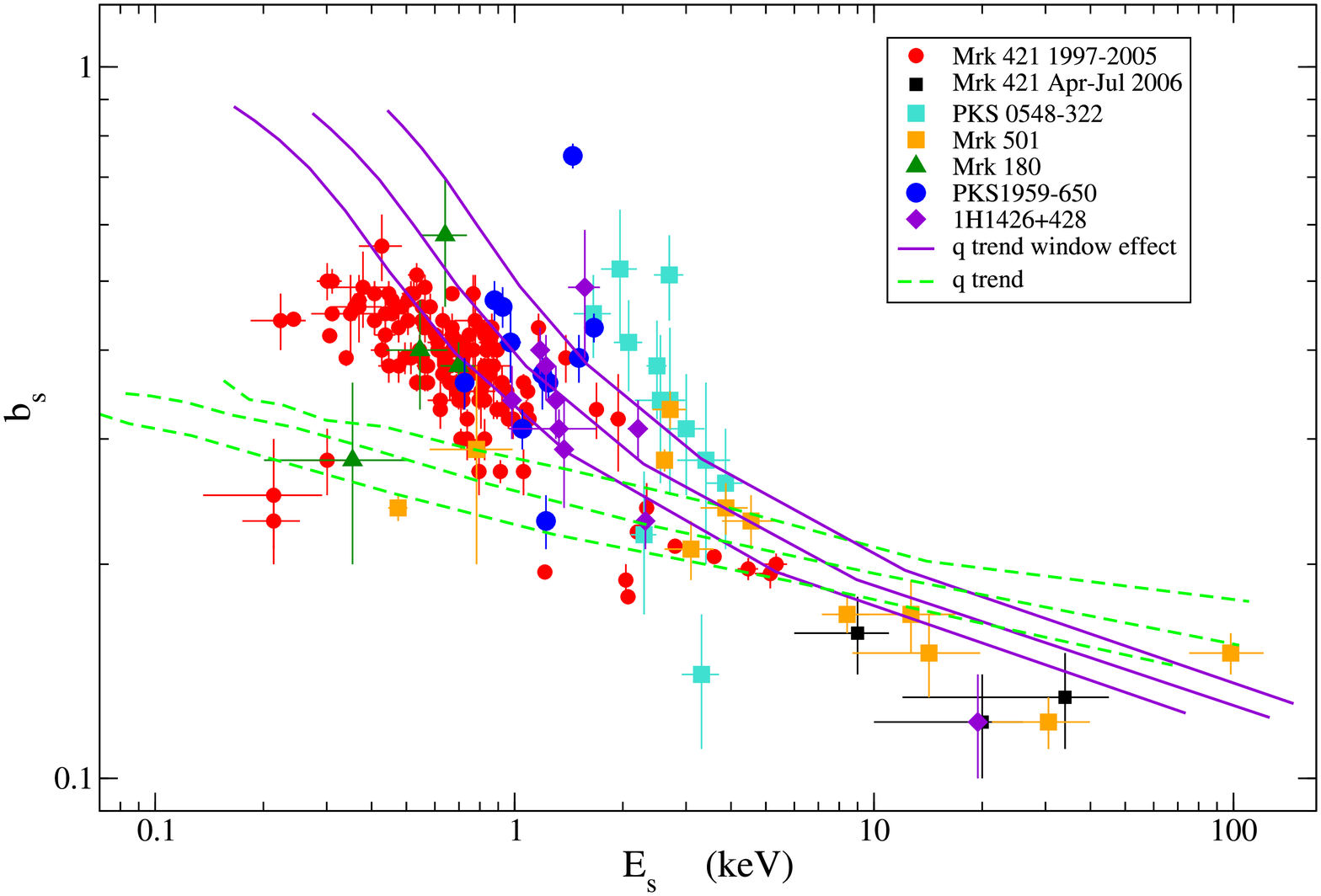}\\
\end{tabular}
\caption{ 
\textit{Left panel:} 
the $E_s$-$b_s$ trend observed for the six HBLs in our sample.
The dashed green lines represent the trend reproduced by stochastic acceleration
model, for the parameters reported in Tab. 3, and for the $D$ trend, the 
different lines corresponding to three different values of $B$ reported in Tab.
3. The purple  lines represent the trend obtained by fitting the numerically computed SED over
a fixed spectral window in the range $0.5-100$ keV.
\textit{Right panel:}
the same as in the left panel for the case of the $q$ trend.
}
\label{fig:fit_data_Es-bs}
\end{figure*}

\setcounter{table}{2}
\begin{table}[ht]
\begin{center}
\caption{Parameters' values adopted in the numerical solutions of the diffusion equation
to reproduce the observed trends of the HBLs reported in Sec. 6}
\begin{tabular}{ll|l|l}
\hline
&&$D$ trend&$q$ trend\\
\hline
\hline
$R$             &(cm)         &$3\times10^{15}$&-      \\
$B$             &(G)          &[0.05-0.2]    &  - \\
$L_{inj}$   ($E_s$-$b_s$ trend)    & (erg/s)    &$5\times10^{39}$&  -  \\
$L_{inj}$   ($E_s$-$L_s$ trend)   & (erg/s)     &$5\times10^{38}, 5\times10^{39}$&  -  \\
$q$             &             &2               &  [3/2-2] \\
$t_A$           & (s)       &$1.2\times 10^3$ & -\\
$t_{D_0 }=1/D_{P0}$  & ($s$)       &[$1.5\times 10^4 -1.5\times 10^5]$ &$1.5\times 10^4$ \\
$T_{inj}$       & ($s$)       &$10^4$            &  -\\
$T_{esc}$       & ($R/c$)     &$2.0 $         &  -    \\
Duration        & ($s$)       &$10^4$         &  -    \\
$\gamma_{inj}$  &             &10.0            &  -    \\
\hline
\end{tabular}
\end{center}
\label{tab:fit_par}
\end{table}

\subsection{ $E_s$-$b_s$ relation}
\label{sec:Es_bs_fit}
The $E_s-b_s$ trend, and in particular the anticorrelation  between these two
observables parameters, is the strongest signature of a stochastic  component in
the acceleration. 

In Fig. \ref{fig:fit_data_Es-bs} we report the scatter plot in the $E_s$-$b_s$ plane
for the six considered sources. 
The left panel reports the results obtained by changing the value of $D_{p0}$:
the green dashed lines describe the trend resulting from a log-parabolic fit of the 
synchrotron SED over a decade in energy centered on $E_s$; the purple lines represent 
the same trend obtained by fitting  log-parabola in the fixed spectral window 
[0.5, 100.0] keV. 
Both these trends are compatible with the data and track the predicted anticorrelation 
between $E_s$ and $b_s$.
Purple data, however, give a better description, hinting that the ``window" effect could be a
real bias. 
Each of the three lines was computed for a different value of the magnetic field.
It is remarkable that the variation of a single parameter, $D_{p0}$ can describe the
observed behaviour. 
The dispersion in the data is relevant, and can be related to the variation of $B$ 
(as partially recovered by numerical computation), or by different values of the beaming
factor, $R$, and $L_{inj}$, during different flares, and for different objects.

The dot-dashed tick line represents the best fit of the observed data by means of
Eq. \ref{eq:Ep_b}, and returns a value of $a\approx 0.6$, as expected from 
theoretical predictions for the case of the $\delta-$approximation, and pure log-parabolic
electron distribution. This fitted line is also compatible with the numerical
trend shown by the purple lines.
Note that the observed curvature values are in the range [0.1, 0.5], corresponding 
to $r_{3p}\sim$ [0.5, 3.0]. According to the results presented in
Sec. \ref{sec:imp_inj}, the expected equilibrium curvature in the synchrotron emission, 
in the full KN or TH regime, and for $q=2$, should be of $r_{3p}\approx 6.0$, and of
$r_{3p}\approx 5.0$ in the intermediate regime. In the case of $q=3/2$, the equilibrium
curvature should be $r_{3p}\sim3.0$.
This is perhaps  an interesting hint  that, both in the flaring and the quiescent states, 
for $q=2$, the distribution is always far from equilibrium. 
In the case of $q=3/2$, only for $E_s\lesssim1.5$ keV the curvature are compatible with 
the equilibrium ($r_{3p}\simeq 3.0$, corresponding to $b_s\sim 0.6$).
For larger values of $E_s$, we find again curvature well below the equilibrium value.
These results provide a good constraint on the values of the magnetic field $B\lesssim 0.1$ G.

The $q$-driven trend (right panel) is also compatible with the data, but for values 
of $E_s\lesssim$ 1 keV, the $D_{p0}$-driven case seems to describe better the observed
behaviour, but any firm conclusion is not possible because of the dispersion of the data.

\begin{figure*}
\centering
\begin{tabular}{cc}
\includegraphics[width=8.5cm,angle=-0]{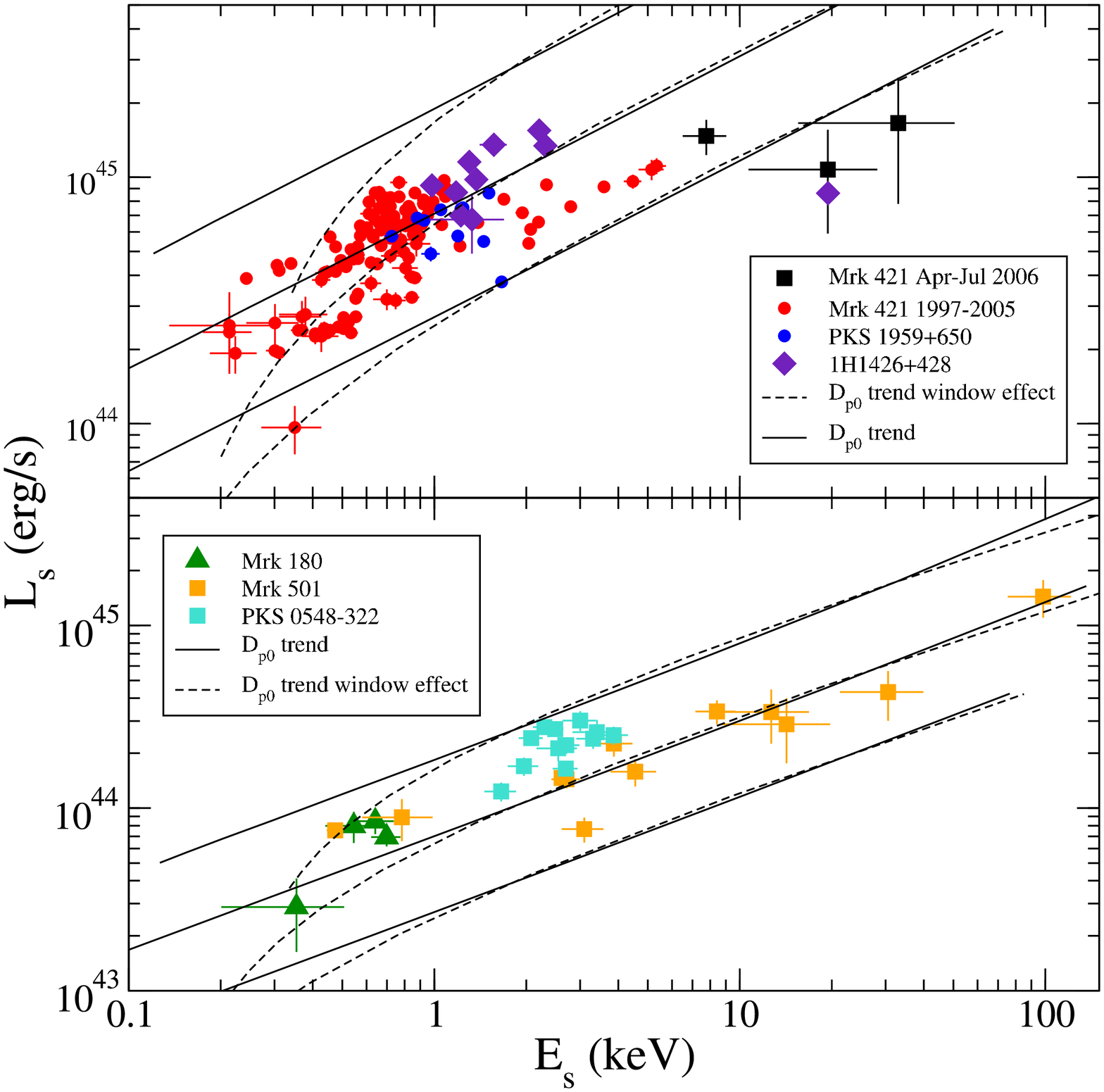}&
\includegraphics[width=8.5cm,angle=-0]{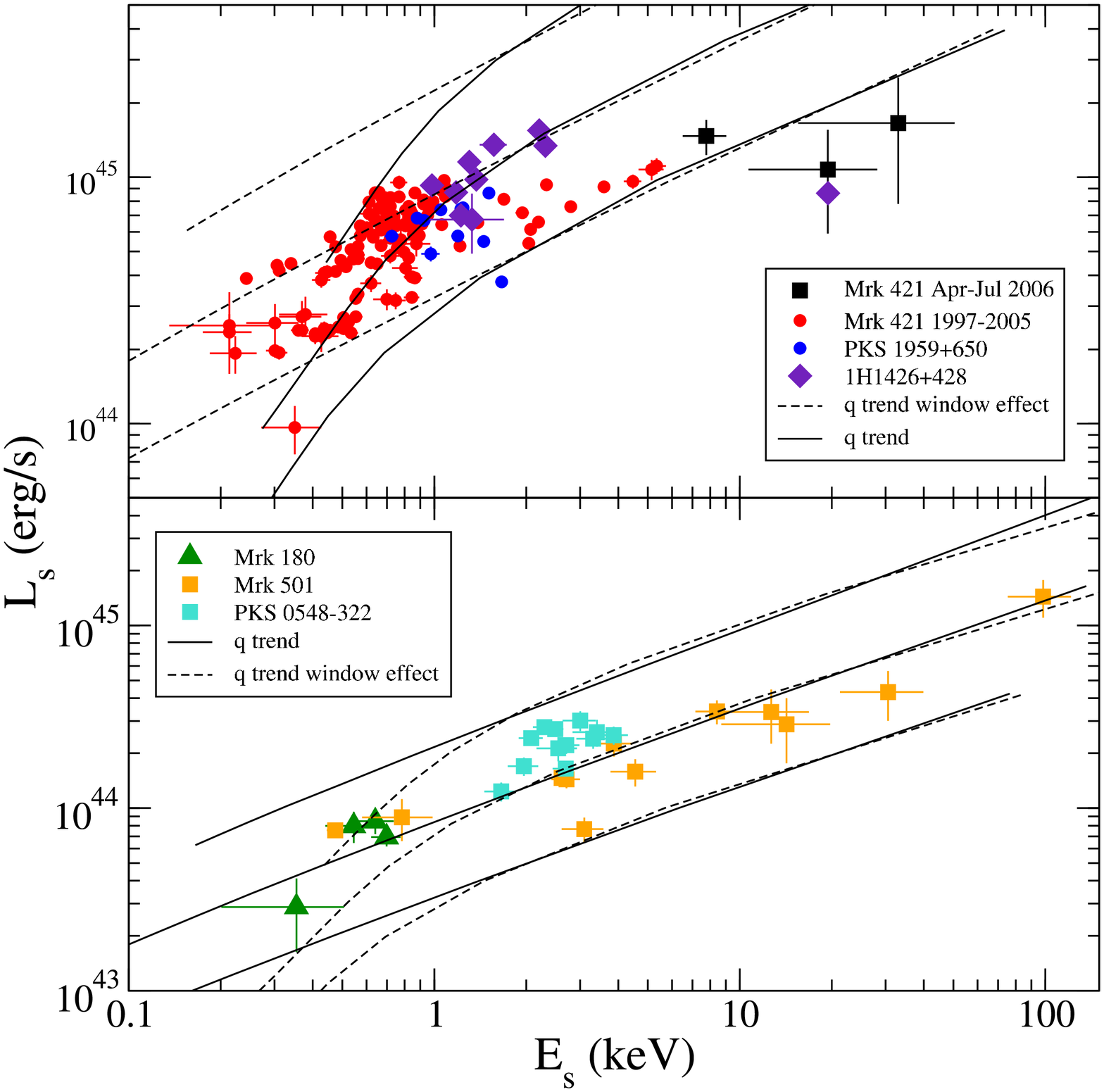}\\
\end{tabular}
\caption{ 
\textit{Left panels:} 
the $E_s$-$L_s$ trend observed for six HBLs in our sample, 
top panel corresponds to the case of $L_{inj}=5\times 10^{39}$ erg/s, bottom
panel corresponds to the case of $L_{inj}=5\times 10^{38}$. 
The solid black lines represent the trend reproduced by stochastic acceleration
model, for the parameters reported in Tab. 3, and for the $D$ trend, the 
different lines corresponding to three different values of $B$ reported in Tab.
3. The dashed  lines represent the trend obtained by fitting the numerically computed SED over
a fixed spectral window in the range $0.5-100$ keV.
\textit{Right panels:}
the same as in the left panel for the case of the $q$ trend.
}
\label{fig:fit_data_Es-Ss}
\end{figure*}

\subsection{ $E_s$-$L_s$ trend}
\label{sec:Es_Ss_fit}
As a last benchmark for the stochastic acceleration model, we reproduce the
observed correlation between $E_s$ and $S_s$, which follows naturally from the 
variations of $D_{p0}$ and $q$.
Considering that the redshifts of the six considered HBL objects are different, 
we prefer to use their peak luminosity $L_s=S_s~4\pi~D_L^2$, where $D_L$ is the luminosity 
distance \footnote{We used a flat cosmology model with:
$H_0$= 73 km/s/Mpc,~~$\Omega_{\rm matter}$=0.27,~~$\Omega_{\rm vacuum}$ = 0.73}.  
To account for the different jet power of sources, we considered 
two data subsets, and we assumed $L_{inj}=5\times 10^{39}$ erg/s  for the first 
subset (top panels of Fig. \ref{fig:fit_data_Es-Ss}), and $L_{inj}=5\times 10^{38}$ 
for second (bottom panels of Fig. \ref{fig:fit_data_Es-Ss}). 
In the left panels of Fig. \ref{fig:fit_data_Es-Ss}, we report
the $D_{p0}$ driven trend, and in the right panels the $q$ driven trend.
Solid lines represent the trend obtained by deriving $L_s$ from
the log-parabolic best fit of the numerically computed SEDs, centered on $E_s$;
dashed lines are the trends obtained by fitting the numerical results in the fixed 
energy window [0.5, 100] keV.

Both the results give a good description of the observed data,
and their shapes are similar. 
Solid lines follow well a power-law with an exponent of about 0.6, while the windowed 
trends (dashed lines) show a break around 1 keV and the exponent below this energy turns to
about 1.5.
A similar break at the same energy, can be noticed in the points of Mrk 421 in 
the $E_s$-$S_s$ plot presented by \cite{Tramacere2009}, who found an exponent of $\sim$ 1.1 
and of $\sim$0.4 below and above 1 keV, respectively. 
This could again be an indication that the observed values are actually affected by the bias.

\section{Discussion}

Broad band observations of non-thermal sources have shown that the spectral
curvature at the peaks of their SEDs can now be measured with  good accuracy.
In this paper, we have presented, using different approaches, the relevance of 
these data for the understanding of the competition between statistical acceleration 
and radiation losses.
First, using a simple statistical approach and Monte Carlo calculations, we have shown
that the log-parabolic energy distribution of the relativistic electron is a
good picture in the first phases before  equilibrium is reached.
In this case the curvature decreases with time and, therefore, for increasing peak
energies. 
This evolution is confirmed by numerical solutions of the diffusion
equation taking properly into account both stochastich acceleration and radiative
SSC cooling. 
The major results can be summarised as follows.

The evolution of the electron energy distributions (Sec. \ref{sec:numeric}) shows that:
\begin{itemize} 
\item
in the case of synchrotron and SSC cooling, and for
all the values of $B$ and $R$, as long as the distribution is far from 
equilibrium, the trend on $r$ is dictated by $D_p$, and is well described by Eq. 
\ref{eq:analyt_sol_curv};
\item 
when the distributions approach  equilibrium, the value of $r$ is determined
by the shape of the equilibrium distribution, which is a relativistic Maxwellian, with 
the sharpness of the cuf-off determined by both $q$ and the IC cooling regime;
\item 
in the case of $q=2$, and for equilibrium energies implying that  IC cooling happens either in the TH regime 
or in the extreme KN regime (IC cooling negligible compared to the synchrotron 
one), the numerical solution of the diffusion equation follows the analytical prediction 
($f=1$, that holds for any $\dot\gamma\propto\gamma^2$), and the corresponding 
equilibrium curvature is $r_{3p}\approx 6.0$ ($b_s\approx 1.2$).
In the case of $q=3/2$ the equilibrium curvature is $r_{3p}\approx 3.0$  ($b_s\approx 0.6$).
These limiting values could be a useful observational test to find cooling dominated flares with the 
distribution approaching to the equilibrium; 
\item 
when  cooling is in the intermediate regime between TH and KN and for the $q=2$ case, 
the condition $f=1$  
fails, and the end values of $r$ decrease, strongly depending on the balance between 
$U_B$ and the seed IC photon energy ($U_{ph}$); numerical computations are necessary 
to evaluate the right value of $r$ at equilibrium. 
\end{itemize} 

The analysis of the spectral evolution of SSC emission (Sec.\ref{sec:ssc_evolution})
shows that:
\begin{itemize} 
\item 
changes of $D_{p0}$ (or $q$) imply that the curvature and peak energy of the 
synchrotron emission are anticorrelated; the $E_s$-$b_s$ trend can be phenomenologically 
described by Eq. \ref{eq:Ep_b};
\item  
The $E_c$-$b_c$ trend presents a clear signature of the transition from
the TH to the KN regime. In particular when the IC scattering approaches the KN
regime we observe a sharp change in the $b_c$, with a positive correlation with $E_c$, 
whilst in the TH regime the correlation is negative as in the case of the $E_s$-$b_c$; 
\item  
the magnetic field plays a relevant role on the cooling process, and $B$ driven 
variations present relevant differences compared to those due to $D_{p0}$ (and $q$).
\end{itemize} 

In particular, for what concerns the B driven case, we note first that the $E_s$-$S_s$
correlation follows the prediction of the synchrotron theory and shows the power-law 
relationship with $E_s\propto(S_s)^{\sim2.0}$.
On the contrary, in the case of $D_{p0}$ and $q$ changes, we find $E_s\propto (S_s)^{0.6}$. 
Another relevant difference in the $B$ driven case is the evolution of $S_c$.  
For the case of $D_{p0}$ and $q$ driven trends $S_c$ relates to $E_c$ through a power-law with exponent 
of about [0.7-0.8]. On the contrary, for the $B$ driven case with IC scattering in the full KN
regime, the value of $E_c$ is almost constant and uncorrelated with $S_c$ (see
Fig. \ref{fig:Ep_vs_b_B_trend}), due to the kinematic limit of the KN regime.
$E_c$ starts to decrease when $B$ is enough large to make dominant the
cooling process. 
This is an interesting signature that could be easily checked in the observed data.

The comparison of the $E_s$-$b_s$ and $E_s$-$S_s$ trends, obtained through several X-ray
observations of six HBL objects spanning a period of many years, with those predicted 
by the stochastic acceleration model, shows very good agreement.
We are able to reproduce these long-term behaviours, by changing the value of 
only one parameter ($D_{p0}$ or $q$). 
Interestingly, the $E_s$-$S_s$ relation follows naturally from that between $E_s$
and $b_s$.
This result is quite robust and hints at a common accelerative scenario 
acting in the jets of HBLs.

As a last remark, we note that very recently \cite{Massaro2011} find
also in the case of GRBs a $E_s$-$b_s$ trend, similar to that observed in
the case of HBL objects. 
They measured values of the curvature up to 1.0, typically higher than in
HBLs. 
It's interesting to note that the value of 1.0 is close to the limit of 
$\sim$1.2, that we predict in the case of distributions approaching the equilibrium 
in either TH or KN regime, for $q=2$.
\\

\acknowledgments
We would like to thank the anonymous referee for providing us with 
constructive comments and suggestions.

This work has been partially supported by Universit\`a di Roma La
Sapienza (Dipartimento di Fisica, Gruppo SCAE).

\bibliography{stoch} 

\end{document}